\RequirePackage{fix-cm}
 
\documentclass[pdflatex,sn-mathphys-num]{sn-jnl}

\usepackage{array}
\usepackage{tabularx}
\usepackage[english]{babel}
\usepackage{setspace} 
\usepackage{amsmath} 
\usepackage{amssymb}
\usepackage{amsfonts}
\usepackage{ulem} 
\usepackage{amsthm}
\usepackage{algorithm}
\usepackage{algorithmic}
\usepackage{xcolor}
\usepackage{tikz}
\usepackage{tabularx}
\usepackage{multirow}
\usepackage{multicol}
\usepackage{makecell}
\usepackage{booktabs}
\usepackage{amsmath}
\usepackage{subcaption}  
\usepackage{graphicx}
\usepackage{graphicx}
\usepackage[utf8]{inputenc}
\usepackage{hyperref}
\usepackage{url}
\usepackage{tikz}
\usetikzlibrary{trees}
\usepackage{epstopdf} 
\newcolumntype{C}[1]{>{\centering\arraybackslash}p{#1}}
\usepackage{geometry}
\theoremstyle{thmstyleone}%
\theoremstyle{thmstyletwo}%

\theoremstyle{thmstylethree}%

\begin{document}

\title[Article Title]{Exploration of  Hepatitis B Virus Infection Dynamics through Physics-Informed Deep Learning Approach}


\author[1]{\fnm{Bikram  } \sur{Das}}\email{bikram.das@iitg.ac.in; rupchandsutradhar@gmail.com}
\equalcont{These authors contributed equally to this work.}

\author*[1]{\fnm{Rupchand } \sur{Sutradhar}}\email{rupchandsutradhar@gmail.com}
\equalcont{These authors contributed equally to this work.}

\author[1]{\fnm{ D C } \sur{Dalal}}\email{durga@iitg.ac.in}
\equalcont{These authors contributed equally to this work.}

\affil[1]{\orgdiv{Department of Mathematics}, \orgname{Indian Institute of Technology
 Guwahati}, \orgaddress{\street{} \city{Guwahati}, \postcode{781039}, \state{Assam}, \country{India}}}



\abstract{Accurate forecasting of viral disease outbreaks is crucial for guiding public health responses and preventing widespread loss of life. In recent years, Physics-Informed Neural Networks (PINNs) have emerged as a promising framework that can capture the intricate dynamics of viral infection and reliably predict its future progression. However, despite notable advances, the application of PINNs in disease modeling remains limited. Standard PINNs are effective in simulating disease dynamics through forward modeling but often face challenges in estimating key biological parameters from sparse or noisy experimental data when applied in an inverse framework. To overcome these limitations, a recent extension known as \textit{Disease Informed Neural Networks (DINNs)} has emerged, offering a more robust approach to parameter estimation tasks. In this work, we apply this \textit{DINNs} technique on a recently proposed hepatitis B virus (HBV) infection dynamics model to predict infection transmission within the liver. This model consists of four compartments: uninfected and infected hepatocytes, rcDNA-containing capsids, and free viruses. Leveraging the power of \textit{DINNs}, we study the impacts of (i) variations in parameter range, (ii) experimental noise in data, (iii) sample sizes, (iv) network architecture and (v) learning rate. We employ this methodology in experimental data collected from nine HBV-infected chimpanzees and observe that it reliably estimates the model parameters. \textit{DINNs} can capture infection dynamics and predict their future progression even when data of some compartments of the system are missing. Additionally, it identifies the influential model parameters that determine whether the HBV infection is cleared or persists within the host. }




\maketitle
\section{Introduction}

Machine learning (ML), one of the tools of data analysis, allows systems to learn from data and perform tasks without being explicitly programmed. ML has multiple applications in diverse fields, such as natural language processing \cite{2020_mankolli_machine}, image processing \cite{2022_li_research}, healthcare \cite{2018_shailaja_machine}, bioinformatics \cite{2017_min_deep}, mathematical biology \cite{2022_greener_guide}, fluid dynamics \cite{2020_brunton_machine}. The connections between mathematical biology and ML have a rich and intricate history \cite{2007_tarca_machine}. ML contributes to understand the dynamics of complex biological systems through data-driven pattern recognition \cite{2020_djellali_data}.
 ML algorithms are capable of addressing some limitations of traditional methods, including handling large datasets \cite{2017_l_machine}, improving the accuracy of prediction \cite{2020_lopatkin_predictive}, and establishing complex and non-linear relationships between model variables and parameters \cite{2020_medina_development}.
 These characteristics make the ML algorithms suitable for forecasting the possible outcomes of infectious diseases that often depend on multiple factors, such as population demographics, environmental conditions,  individual behaviors. In case of mathematical biology, accurate parameter estimation is utmost important to understand the disease dynamics and make reliable predictions of outcomes. Inaccurate parameter estimation prevents the models from reflecting the actual behavior of the system and may misguide in proposing new therapeutic strategy.
 Several statistical methods, such as linear and non-linear least-squares fitting technique \cite{1998_mendes_non}, optimal control strategy \cite{2018_tsiantis_optimality}, Bayesian framework \cite{2007_wilkinson_bayesian}, particle swarm optimization \cite{2018_mosayebi_modified}, agent-based modeling \cite{2010_dancik_parameter}, maximum likelihood estimation \cite{2004_muller_tests}, Bayesian inference \cite{2006_huang_hierarchical}, and Poisson regression \cite{1985_frome_use}, were introduced to estimate the model parameters 
 using real-world data.
 Although these approaches show significant contributions for comparatively large datasets,  the demands for more efficient computational methods for parameter estimation remain high in case of tiny or limited datasets. Neural networks, especially Physics-Informed Neural Networks (PINNs), have emerged as a universal function approximator and are capable of solving complex, non-linear system of differential equations \cite{2019_raissiphysics,2018_raissihidden}, such as those seen in biological processes. PINNs are now widely applied in various fields, including computer vision \cite{2024_banerjee_physics}, natural language processing \cite{2020_guo_solving}, complex fluid dynamics \cite{2024_zhao_comprehensive}, medical science \cite{2022_sarabian_physics} and more recently to the modeling of infectious diseases \cite{2019_raissiparameter}. By combining the concepts of ML and the underlying principles of physics, PINNs present an innovative approach to address intricate problems governed by the laws of physics.  Due to the advantages of PINNs, numerous studies in the literature have applied this approach to analyze various infectious diseases. In epidemiology, the application of PINNs is rapidly expanding, particularly in forecasting the outbreaks of Covid-19 \cite{2024_han_approaching}, tuberculosis \cite{2025_danumjaya_dynamic}, dengue \cite{2024_viet_adapting}, and other diseases. However, their application to Hepatitis B, particularly in modeling within-host viral dynamics, remains limited, making this study novel and timely.
 
Hepatitis B virus (HBV) is a major public health concern worldwide, leading to significant morbidity and mortality. According to  World Health Organization, approximately 254 million people are living with chronic HBV infection throughout the world. Despite this high prevalence, only 7 million cases with chronic hepatitis B were diagnosed \cite{WHO_2024}. Although this viral infection is preventable with safe, widely available, and effective prophylactic vaccines, it remains a significant global issue. This viral infection can be  acute or chronic.
Acute hepatitis B lasts less than six months and is generally eliminated through the immune response, with clearance rates ranging from 85\% to 95\%  \cite{2001_whalleykinetics,2007_ciuperole}.
 On the other hand, chronic HBV infection (CHBV) is a lifelong and incurable condition of the patients, poses serious health risks and emotional challenges, such as stigma, anxiety, and the financial burden of sustained healthcare \cite{2002_ribeirodynamics}. If CHBV infection is kept untreated, it can result  serious liver disorders, such as cirrhosis, hepatocellular carcinoma (HCC), and  other potentially life-threatening complications. At present, pegylated interferons (immune modulators) and nucleos(t)ide analogues (lamivudine, adefovir, telbivudine, entecavir, and tenofovir) serve as  effective  therapeutic options for  CHBV infection \cite{2018_terrault_update}.

Mathematical models provide valuable insights for understanding and investigating the spread and control of viral infections. Especially to HBV infection, numerous mathematical models have been developed to explain the dynamics of this disease. In 1996, Nowak et al. \cite{1996_nowak_viral} introduced a pioneering mathematical model on HBV infection dynamics and it is still considered as the `basic model' in the field of viral dynamics. 
Following  this basic model \cite{1996_nowak_viral}, extensive research \cite{2006_murrayhalf,2007_ciupe_modeling,2008_min_mathematical,2021_hews_global,2023_sutradhar_fractional} has been carried out, either by modifying it or by proposing new models to capture the complexities of this infection dynamics in a better way. 
Min et al. \cite{2008_min_mathematical} modified the basic model (mentioned in \cite{1996_nowak_viral})  by substituting the mass action term with a standard incidence function.  
By formulating another  HBV dynamics model considering infected hepatocytes, capsids and viruses, Murray et al. \cite{2006_murrayhalf} estimated the HBV half-life to be approximately of 4 hours. Fatehi et al. \cite{2020_fatehi_intracellular} developed an intracellular model on this viral infection and  discussed various therapeutic strategies that could be applied in the future. By fitting human acute infection data along with physiological constraints to five mathematical models and comparing  Akaike Information Criterion ($\text{AIC}_\text{c}$), Goyal et al. \cite{2017_goyal_role} concluded that without causing lethal loss of liver mass, the clearance of acute HBV infection is strongly associated with the cellular proliferation of  infected hepatocytes resulting in two uninfected progenies. Beyond the studies mentioned above, a broad range of literature exists that examined various aspects of HBV infection. Nevertheless, none of the aforementioned studies examined the roles of capsid recycling in the infection.
Capsid recycling refers to the process in which a portion of newly produced rcDNA-containing capsids is transported back to the nucleus, increasing the amount of super-coiled covalently closed circular DNA. This mechanism plays a crucial role in maintaining a consistent reservoir of cccDNAs and contributes significantly to the replication of viruses. Recently, Sutradhar and Dalal \cite{r2024_sutradhacytoplasmic} proposed  the following model (with some symbolic modification) on this viral infection by incorporating the cytoplasmic recycling of rcDNA-containing capsids: 
\begin{equation}
   \label{main model}
  \left.
  \begin{aligned}
  \frac{dX}{dt} &= \lambda - \mu X - kVX,\\
  \frac{dY}{dt} &= kVX - \delta Y, \\
  \frac{dD}{dt} &= aY + \gamma(1 - \alpha)D-\alpha \beta D -\delta D,\\
  \frac{dV}{dt} &= \alpha \beta D - cV.
  \end{aligned}
  \right\}
\end{equation}

\noindent Here, dependent variables $X,Y,D$ and $V$ represent the concentrations of susceptible hepatocytes, infected hepatocytes, rcDNA-containing capsids, and free viruses, respectively. The values of all model parameters $\lambda, \mu, k, a, \delta, \alpha, \beta, \gamma$ and $c$ are non-negative real numbers and their biological interpretations are given below:
\begin{itemize}
    \item $\lambda$: Natural growth rate of uninfected hepatocytes.
    \item $\mu, \delta$: Per capita death rates of uninfected and infected hepatocyte. 
    \item $k$: Disease transmission rate constant.
    \item $a$:  Production rate of  rcDNA-containing capsids per infected hepatocyte.
    \item $\gamma$: Recycling rate of newly produced capsids within the cytoplasm of infected hepatocyte.
    \item $\alpha$: Volume fraction of newly produced capsids  that contributes to viral production.
    \item $\beta$: The rate of export of rcDNA-containing capsids  to the blood as new viruses. 
    \item $c$: Clearance rate of viruses. 
\end{itemize}

\noindent Due to the incorporation of capsid recycling, it is observed that the model \eqref{main model} reveals rich dynamics of HBV infection, along with several key findings. However, in order to validate most of the mathematical models, including \eqref{main model}, authors generally use the traditional numerical methods. These methods often have deficiencies in handling tiny datasets, which are commonly encountered in mathematical biology. In such cases, the machine learning approaches, especially PINNs, may prove to be more effective and useful.

Recent advancements in PINNs have demonstrated their potential in learning infectious disease dynamics and estimating model parameters. However, it is observed that their applications to parameter estimation are limited to those compartmental models (e.g., SIR model) having small differences in order of magnitude between any two parameters. In real-world scenarios, viral infections are often governed by a set of parameters that have significant variation in the order of magnitude relative to each other. For example, in the study of Sutradhar and Dalal \cite{r2024_sutradhacytoplasmic}, one possible set of values for the parameters $\lambda$ and $k$ was considered as $2.6\times 10^{7}$ and $1.67\times 10^{-13}$. These two parameters differ by approximately 20 orders of magnitude, \textit{i.e.}, $\frac{o(\lambda)}{o(k)} \approx 20$. In such cases, standard PINNs often struggle to estimate parameter values from the available experimental data, and consequently fail to reliably capture the underlying infection dynamics.
In order to address these challenges, a novel approach known as \textit{Disease Informed Neural Networks (DINNs)} \cite{2022_shaier_data} has emerged, offering a more robust technique for parameter estimation tasks. The key idea behind this novel approach lies in
\begin{enumerate}
\item Designing a neural network that ensures normalization of the input data.

\item Restricting the unknown parameters to specific ranges so that the model can effectively learn the parameters with vastly different orders of magnitudes from the experimental data.

\end{enumerate}

In this study, we apply this \textit{DINNs} approach to a recently proposed HBV infection dynamics model. 
By employing the inverse solver strategies, this approach holds strong capability in estimating all the parameters of the system \eqref{main model} from the experimental data.
By leveraging the versatility of \textit{DINNs}, we investigate the effects of the following factors on HBV dynamics: (i) variations in parameter range, (ii) experimental noise in data,  (iii) data sample sizes,  (iv) network architecture, and (v) learning rate. Despite the unavailability of the experimental data of uninfected hepatocytes, infected hepatocytes, and viruses, it is shown that \textit{DINNs} can learn the dynamics of infection using only the available experimental data of HBV DNA-containing capsids from nine young, healthy, HBV-seronegative chimpanzees. This framework is also crucial in identifying
the peaks of the infection that can aid in optimizing the timing of antiviral interventions. Moreover, \textit{DINNs} are highly effective in pinpointing the most significant parameters influencing HBV dynamics, which may determine whether the infection is cleared or progresses to a chronic state within the host. Utilizing the estimated values of parameters, we show that this approach effectively forecast the concentration of HBV DNA-containing capsids, highlighting its potential as a powerful tool for predicting infection dynamics.

\section{Equilibrium points and basic reproduction number}
\noindent
The system of ODEs \eqref{main model} has two  equilibrium points: (i) disease-free, and (ii) endemic. In terms of biology, the  feasibility of endemic equilibrium point depends on the value of basic reproduction number $\mathcal{R}_0$.   The disease-free equilibrium point, denoted by \( E_d \), is given by  $E_d = \left( \dfrac{\lambda}{\mu},\ 0,\ 0,\ 0 \right)$. On the other hand, the endemic equilibrium point can be expressed  as
$E_e = \left( 
\dfrac{c \delta R_e}{a \alpha \beta k},\ 
\dfrac{a \alpha \beta k \lambda - c \delta \mu R_e}{a \alpha \beta \delta k},\ 
\dfrac{a \alpha \beta k \lambda - c \delta \mu R_e}{\alpha \beta \delta k R_e},\ 
\dfrac{a \alpha \beta k \lambda - c \delta \mu R_e}{c \delta k R_e}
\right),$
$\quad \text{where} \quad R_e= \alpha \beta -(1-\alpha)\gamma + \delta.$
  In the context of viral infection, the basic reproduction number is defined as the expected number of secondary infected cells
 that are produced by a single infected cell
when all the uninfected cells  are considered to be susceptible to infection \cite{2015_martcheva_introduction}.
In order to compute the basic reproduction number, the next-generation matrix approach is  employed \cite{2002_van_reproduction} and it is  given by 
$$\mathcal{R}_0 = \frac{a k \lambda \alpha \beta}{(c \alpha \beta \delta - c \gamma \delta + c \alpha \gamma \delta + c \delta^2)\mu}. $$
\( \mathcal{R}_0 \) plays a crucial role in determining whether an emerging infectious disease can persist within a population. 
The disease-free equilibrium point is globally asymptomatically stable when $\mathcal{R}_0<1$, while the endemic equilibrium point becomes globally asymptotically  stable if $\mathcal{R}_0>1$ and $R_e>0$ \cite{r2024_sutradhacytoplasmic}. The steady-state of $E_c$ represent the chronic situation of the patients.  Chronic hepatitis B (CHB) is much more serious 
than acute
because
CHB causes long-term morbidity,
 leading to significant public health impact.
Therefore, this study focuses on chronic infection and ensures that the values of the model parameters  
maintain the condition \( \mathcal{R}_0 > 1 \) throughout the rest of this study.

\section{\textit{Disease Informed} deep learning method}

\subsection{Feed-forward neural network (FNN)}

\begin{figure}[ht!]
  \centering
  \includegraphics[scale=0.4]{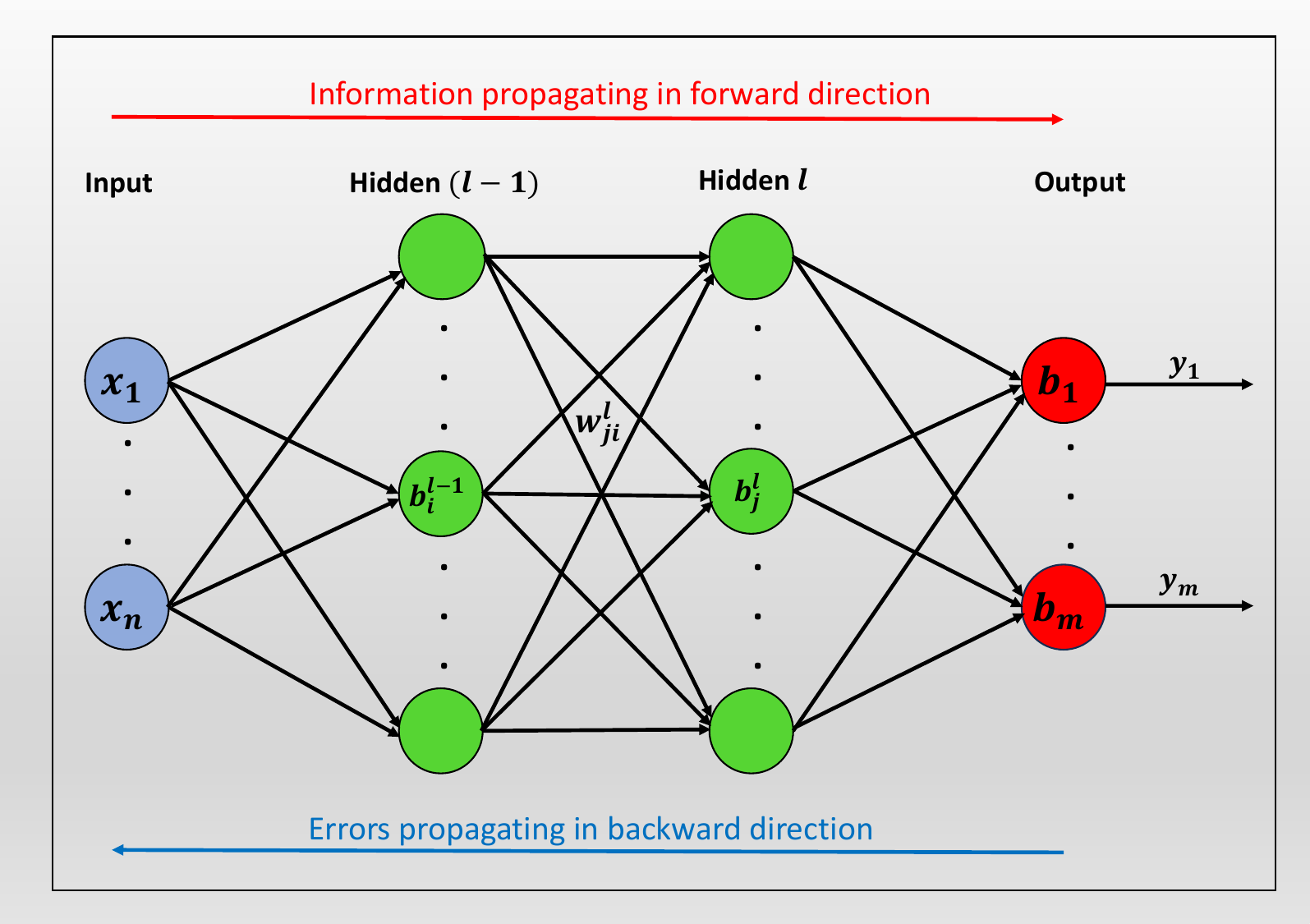} 
  \caption{Architecture of a fully connected feed-forward neural network.}
  \label{fig: Feed-forward Neural Network}
\end{figure}

Generally, a neural network mimics how the human brain functions by using interconnected neurons \cite{2015_lecundeep}. The structure of a fully connected feed-forward neural network, shown in Figure~\ref{fig: Feed-forward Neural Network}, consists of the following layers: (i) an input layer, (ii) one or more hidden layers, and (iii) an output layer. Each layer is made up of units, called neurons, which are represented by colored circles in Figure~\ref{fig: Feed-forward Neural Network}. The neurons in adjacent layers are interconnected and each connection has an associated weight. Specifically, the weight linking the neuron \( i \) in the $(l-1)^{th}$ hidden layer to the neuron \( j \) in the $l^{th}$ hidden layer is denoted by \( w_{ji}^l \), as illustrated in Figure~\ref{fig: Feed-forward Neural Network}. No connections exist between neurons within the same layer or between non-adjacent layers. Input values  \( x_i(i=1,2,...,n) \) propagate through the network via these interconnections, starting at the input layer, passing through the hidden layers, and reaching the output layer which produces output values \( y_j(j=1,2,...,m) \).

The activation function for each neuron's output introduces non-linearity into the neural networks, enabling effective backpropagation by providing gradients that update the weights and biases using the error. The activation function in $l^{th}$ layer  is given as \( \sigma \). In the literature, various activation functions have been proposed to facilitate computation and adaptation in neural networks, such as the sigmoid function, hyperbolic tangent (Tanh), and Rectified Linear Units (ReLU) \cite{2017_sharma_activation}. Recently, the activation functions like Swish, LeCun's Tanh, Bipolar Sigmoid, Mish, and Arctan have  also  been used to improve the  smoothness of the network's output \cite{2017_sharma_activation,2022_wang_smish,2018_zhang_efficient}.  
For effective training, the selected activation functions should be smooth enough to prevent gradient vanishing and ensure stable learning of the patterns. The values at each neuron in the hidden and output layers are computed by summing the weighted outputs from the previous layer and adding a bias. For the $j^{th}$ neuron  in the $l^{th}$ hidden layer, an intermediate quantity \( a_j^l \) is defined as follows:  
\begin{equation}
a_j^l =  b_j^l + \sum_i w_{ji}^l y_i^{l-1},
\end{equation} 
\noindent
where
\( y_i^{l-1} \) represents the output from the previous layer, and \( b_j^l \) indicates the bias corresponding to the $j^{th}$ neuron in the $l^{th}$ layer. Subsequently, the output of the $j^{th}$ neuron in the $l^{th}$ layer is  obtained by applying the activation function $\sigma$ and given as follows: 
\begin{equation}
y_j^l = \sigma \left( a_j^l \right) = \sigma \left( b_j^l + \sum_i w_{ji}^l y_i^{l-1} \right).
\end{equation}

\subsection{Backpropagation Algorithm}
\begin{figure}
  \centering
  \includegraphics[width=0.6\textwidth]{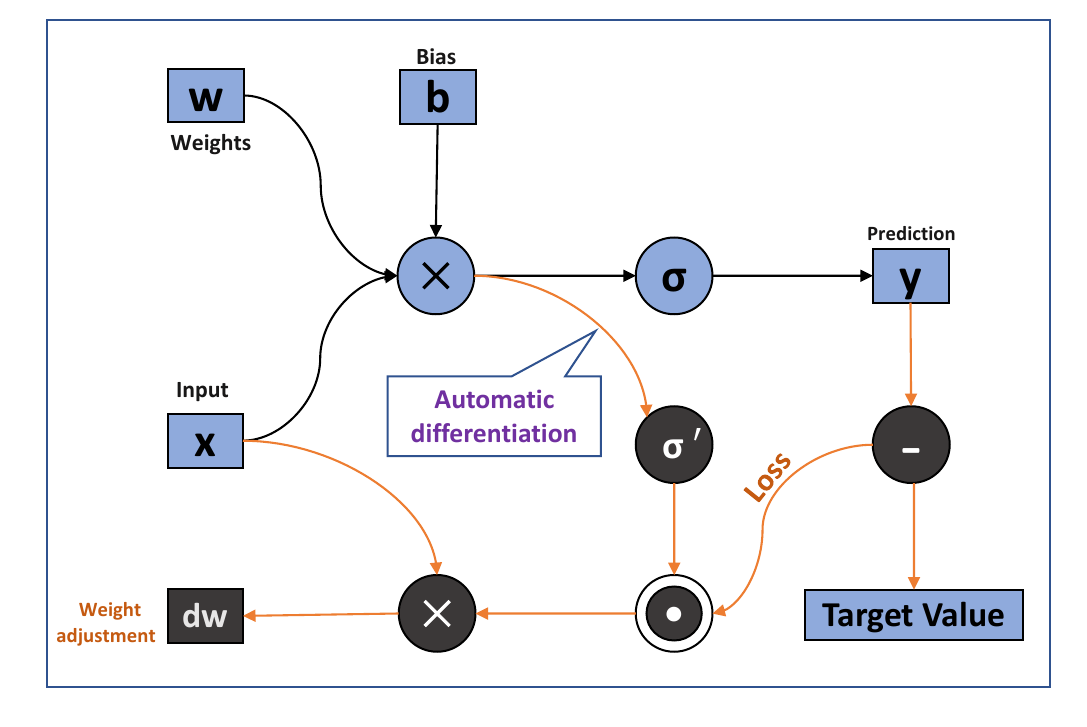} 
  \caption{The schematic representation of backpropagation for FNN.}
  \label{fig: The Schematic of Backpropagation for FNN}
\end{figure}

In order to train the multilayer feed-forward network, the backpropagation (BP) algorithm \cite{2020_lillicrap_backpropagation} is generally applied. This is a supervised learning method in which the outputs of the network are compared to a known target during training to indicate how well the network is performing. In order to optimize the weights and biases of the network, a loss function is considered. The BP algorithm is then used to compute the gradient of the loss function. This process utilizes the chain rule to compute the derivative of the loss function with respect to the weights and biases. The weights and biases are then updated iteratively to minimize the loss, as illustrated in Figure \ref{fig: The Schematic of Backpropagation for FNN}. This process ensures appropriate weight and bias adjustments, enhancing the ability of  network to learn from data.
 The BP algorithm is given as follows: 

\begin{algorithm}
\caption{Backpropagation algorithm}
\begin{algorithmic}
\STATE \textbf{Input:} Generate random sampling points \( x = (x_1, \dots, x_m)^T \) within the domain of interest
\STATE \textbf{Output:} Compute the gradient of the loss function
\STATE \textbf{Initialize:} Set the corresponding activation \( a^1 = x^i, \; i = 1, \dots, m \) for the input layer

\STATE \textbf{Forward Propagation:}\\
~~\text{\quad}{ \textbf{for} $l$ from $2$ to $L$ \textbf{do} Calculate \( z^l = w^l a^{l-1} + b^l \) and \( a^l = \sigma \left( z^l \right) \)}\\
    \quad \quad \textbf{end for}

\STATE \textbf{Output error:} Find output error as \( \delta^L = \nabla_a \text{Loss} \odot \sigma' \left( z^L \right) \)

\STATE \textbf{Backpropagate error:}\\
~~\text{\quad}{ \textbf{for} $l$ from $L-1$ to $2$ \textbf{do} Calculate \( \delta^l = \left( \left( w^{l+1} \right)^T \delta^{l+1} \right) \odot \sigma' \left( z^l \right) \)}\\
\quad \quad \textbf{end for}

\STATE \textbf{Output gradient:} The gradients for the loss function are \( \dfrac{\partial \text{Loss}}{\partial w_{jk}^l} = a_k^{l-1} \delta_j^l \) and \( \dfrac{\partial \text{Loss}}{\partial b_j^l} = \delta_j^l \)

\end{algorithmic}
\end{algorithm}

\subsection{Architecture and methodology of \textit{DINNs}}
\begin{figure}
  \centering
  \includegraphics[width=0.65\textwidth]{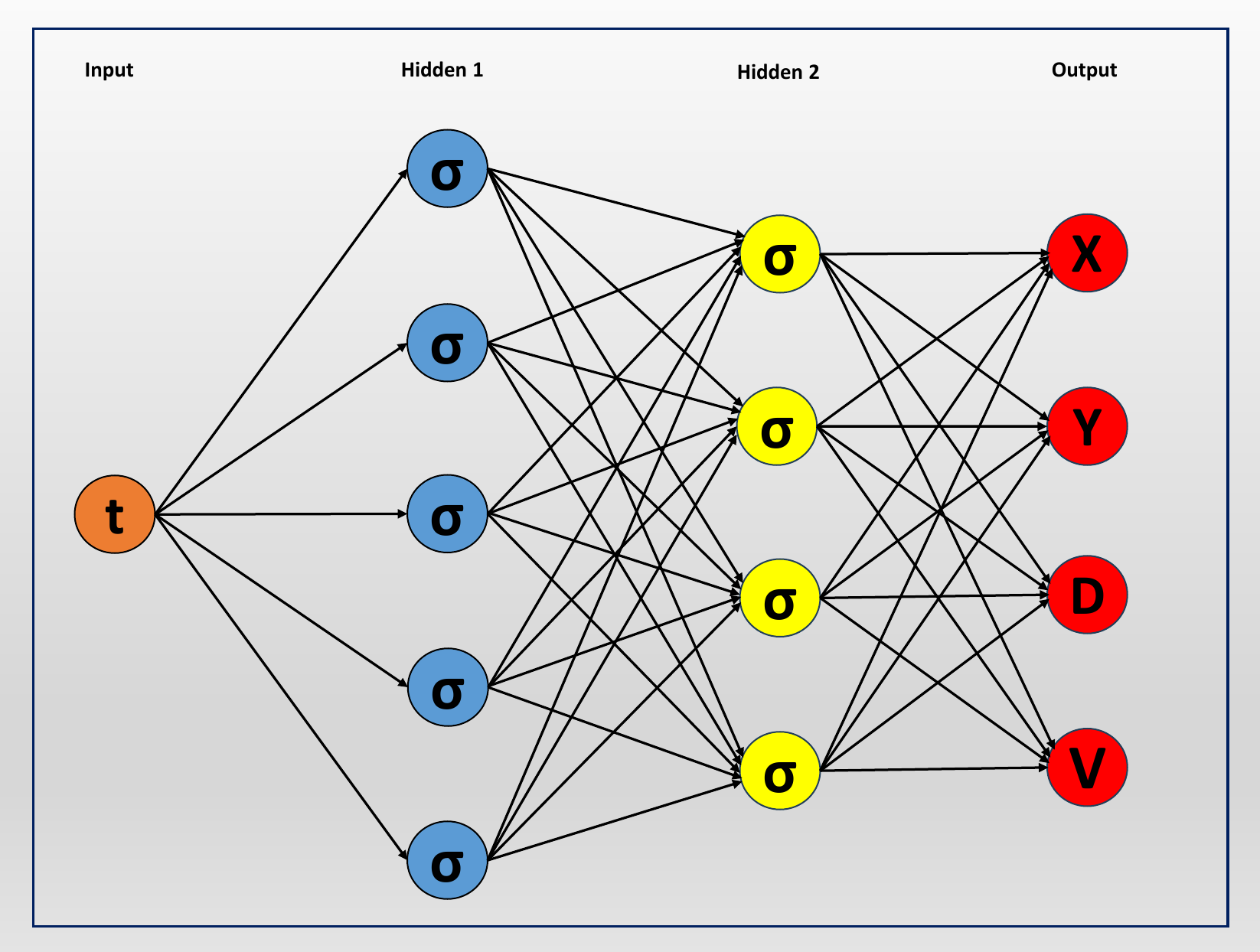} 
  \caption{The  architecture of \textit{DINNs}. The input layer takes the values of the temporal data, and the output layer produces predictions for the dependent variables, \( X \), \( Y \), \( D \) and \( V \).}
  \label{fig: DINNs Architecture}
\end{figure}

Here, we outline the foundational concepts and methodology behind \textit{DINNs}. Figure \ref{fig: DINNs Architecture} presents a structural overview of \textit{DINNs}, illustrating a neural network with two hidden layers. The first hidden layer is composed of five neurons, while the second one consists of four neurons. For a given activation function, the main challenge yields  to determine the optimal weights and biases using backpropagation to best fit the predicted data to the observed values.

\textit{DINNs} primarily aim to harness the underlying dynamics of the system \eqref{main model} and estimate the dependent variables ($X$, $Y$, $D$ and $V$) using deep neural networks. This method employs advanced computational techniques to tackle both forward and inverse problems related to differential equations whereas the unknown functions are approximated using either a neural network or a gaussian process. In this case, the output functions are approximated as
\begin{align}
   t \mapsto (X, Y, D, V) 
\end{align}
via a deep neural network and the corresponding expressions of residuals for the system of equations \eqref{main model} are given as follows:

\begin{equation} \label{residual}
\left.
\begin{aligned}
\displaystyle f_1:	&= \dfrac{dX}{dt} - \lambda+\mu X + kXV, \\ 
\displaystyle f_2:	&= \dfrac{dY}{dt} - kVX + \delta Y, \\ 
\displaystyle f_3:	&= \dfrac{dD}{dt} -  aY - \gamma(1-\alpha)D + \alpha\beta D + \delta D,\\ 
\displaystyle f_4:    &= \dfrac{dV}{dt} - \alpha\beta D + cV. 
\end{aligned}
\right\}
\end{equation} 

 \noindent In order to compute the residuals \( f_1\), \(f_2\), \(f_3\) and \(f_4\), the necessary derivatives are obtained using automatic differentiation \cite{JMLR:v18:17-468}.
 In this computations, a fully connected neural network is implemented by taking the input \(t\) and providing outputs for \( X\), \( Y\), \( D\) and \( V \). The parameters of the system of equations \eqref{main model} are considered as the parameters for the residuals \(f_1\), \(f_2\), \(f_3\) and \(f_4\). The total loss function is defined by combining the regression loss for \( X\), \( Y\), \( D\) and \( V \) and the residual loss enforced by the residuals given in equation \eqref{residual}. The gradients of this total loss function are propagated through the network using a gradient-based optimization algorithm to adjust the parameters of the model \eqref{main model}.

\subsection{Parameter estimation through \textit{DINNs}}

The reliability of an algorithm is validated by its robustness when it is applied to unseen data. For this purpose, model \eqref{main model} is solved in a forward manner by applying the LSODA algorithm to generate a dataset  \cite{2017_tangherlonilassie}. Biologically relevant initial conditions  and the parameter values are taken from the available literature \cite{2006_murrayhalf,2015-murraysilico,r2024_sutradhacytoplasmic,2020_bachraouianalysis}. 
The generated dataset is then fed into the architecture of \textit{DINNs}, ensuring that the input data are normalized. This allows the performance of \textit{DINNs} to be assessed by evaluating how accurately it predicts the true parameter values across a wide range of initial guesses.
This dataset comprises 10 to 500 data points. This dataset is fed into the neural network without providing any prior information about the parameters. Unlike conventional neural network training, this approach does not depend on separate training, validation, and test datasets. Instead, the network learns directly from the data representing the dynamics of HBV spread over time and subsequently tries to predict the value of the parameters that was chosen to generate the data.  Let the unknown solution of \eqref{main model} be
$$u(t;p)= \left[X(t;p),Y(t;p),D(t;p),V(t;p)\right]^{T},$$
where $p$ are the parameters associated to the system \eqref{main model}. Training data is taken as time $\{t^{i}\}$ and solution $\{u^{i}\}$, where $i=0,1,...,N_{u}$ and $N_u$ represents the total number
of data points available for training. $t^{0}$ corresponds to the initial condition for the system \eqref{main model}. The primary focus lies in training the network using parameters $\left(p^{T}, \theta^{T}\right)^{T}$, where $\theta$ denotes the concatenation of weights and biases for all artificial neurons. Later,  a vector $\left(\hat{p}^{T}, \hat{\theta}^{T}\right)^{T}$ and an approximation $\hat{u}^{i}\left(\hat{p}^{T}, \hat{\theta}^{T}\right)$ are obtained using an optimization algorithm.

\noindent The optimization process uses the following loss function
\begin{equation}
\text{MSE} = \text{MSE}_{\text{Net\_HBV}} + \text{MSE}_{\text{Net\_F}},
\label{MSE}
\end{equation}
where
\begin{equation}
\text{MSE}_{\text{Net\_HBV}} =  \sum_{k=1}^{4} \frac{1}{N_u} \sum_{i=1}^{N_u} \left( u^{i}_{k} - \hat{u}^{i}_{k}\left(\hat{p}, \hat{\theta}\right)\right)^2
\end{equation}

and
\begin{equation}
\text{MSE}_{\text{Net\_F}} = \frac{1}{N_{f}} \left(\sum_{k=1}^{N_f} \sum_{j=1}^{4} \left( f^{k}_{j}\left(\hat{p}, \hat{\theta}\right) \right)^2\right).
\label{MSE_NETF}
\end{equation}

\noindent

\noindent In this context, $\text{MSE}_{\text{Net}\_\text{HBV}}$  corresponds to the regression loss between the training data $u^i=(X^{i},Y^{i},D^{i},V^{i})^T$ and the network's prediction $\hat{u}^i=(\hat{X}^{i},\hat{Y}^{i},\hat{D}^{i},\hat{V}^{i})^T$, where $i=1,2,...,N_{u}$. 
On the other hand, $\text{MSE}_{\text{Net}\_\text{F}}$ enforces the loss imposed by the system \eqref{residual} at a finite set of measurement points, $N_f$, whose number and locations are taken to be the same as the training data. It is important to note that the points used to enforce the differential equations might have different numbers and locations compared to the actual training data.

\vspace{0.1cm}
\noindent In order to improve the performance of PINNs, many studies have proposed loss re-weighting strategies based on the training dynamics of PINNs from various perspectives and under different assumptions \cite{2021_wang_understanding,2022_maddu_inverse}. However, a fair and comprehensive benchmark for comparing these methods and selecting optimal loss weights is still lacking. Moreover, such methods typically introduce additional computational complexity and often require extensive hyperparameter tuning. Therefore, we assign an equal weighting scheme $(1:1)$ for the data loss and the residual loss in equation \eqref{MSE}, similar to commonly used in the PINNs literature.
 The algorithm~\ref{DINNs algorithm} outlines the procedure for estimating the parameters $p$ and the solution $u(t; p)$.
\begin{algorithm} 
\caption{Algorithm behind \textit{DINNs}}
\label{DINNs algorithm}
\begin{algorithmic}
\STATE \textbf{Input:} Take training data $\{t^{i}\}$ and $\{u^{i}\}$, where $i=1,2,...,N_{u}$ 
\STATE \textbf{Output:} Compute $\hat{p}$ and $\hat{u}$
\STATE \textbf{Initialize:} Set the initial values of the parameters $\hat{\theta}_{0}$ and $\hat{p}_{0}$
\STATE \textbf{1.} Set the time interval over which the solution will be computed
\STATE \textbf{2.} Set the loss function as a combination of data loss and residual loss
\STATE \textbf{3.} Design an FNN with a single input neuron and four output neurons (each corresponding to a compartment), ensuring that the input data is normalized
\STATE \textbf{4.} Select appropriate optimization hyper-parameters, such as the Adam optimizer and learning rate
\STATE \textbf{5.} {\textbf{for} \textit{iter} from $1$ to $max\_iter$ \textbf{do}:

~~\text{\quad}\textbf{a.} Calculate the total loss $MSE$. It is required to use automatic differentiation for the residuals

~~\text{\quad}\textbf{b.} Train the network using optimizer algorithm and update $\hat{\theta}_{iter-1}$ to $\hat{\theta}_{iter}$

~~\text{\quad}\textbf{c.} Get the values $\hat{p}_{iter}$ and $\hat{u}_{iter}$}

\STATE \textbf{6.} \textbf{return}  $\hat{p}_{\mathrm{max}_{\mathrm{iter}}}$ and $\hat{u}_{\mathrm{max}_{\mathrm{iter}}}$
\end{algorithmic}
\end{algorithm}

\section{Computational experiments with \textit{DINNs}}
\begin{figure}[ht!]
  \centering
  \includegraphics[width=13cm, height=8cm]{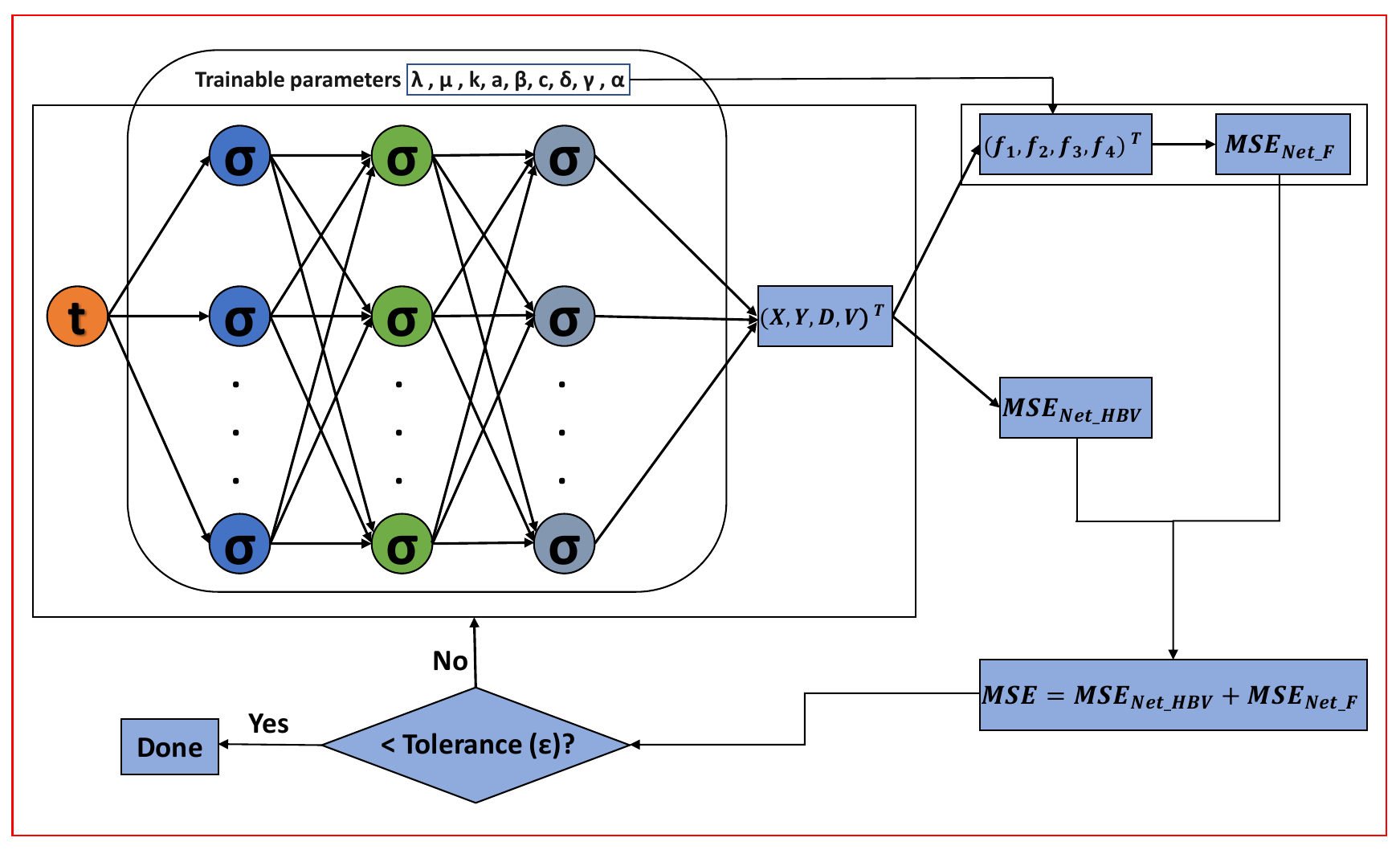}
  \caption{Schematic representation of \textit{DINNs} to learn the dynamic along with model parameters. Here, $\epsilon$ represents the permissible margin of tolerance.}
  \label{fig: Parameter estimation photo}
\end{figure}

In this section, we carry out a series of numerical experiments employing \textit{DINNs} to uncover the dynamics of the infection and to estimate the associated parameters of the model \eqref{main model}. A schematic diagram illustrating the workflow of \textit{DINNs} is shown in Figure \ref{fig: Parameter estimation photo}. The neural network considered in this experiments consists of four fully connected layers. Each hidden layer contains twenty neurons.  The activation function, ReLU, is applied between the layers. 
One of the main benefits of the ReLU activation function, compared to others such as sigmoid or tanh, is that it can prevent all neurons from being activated simultaneously.
The network is trained on an Intel(R) Core(TM) i5-10210U CPU @1.60 GHz. Depending on the complexity of the system, the training duration ranges from 2 to 12 hours, though this duration can be significantly shortened with the use of TPUs and GPUs. 
 Learning both the dynamics of the system  and associated unknown parameters results in a longer training duration. However, if the values of the parameters are provided, the system takes much less time to train itself and captures only
 the dynamics of the model \eqref{main model}. The learning rate is adjusted using PyTorch’s CyclicLR scheduler \cite{2024_shao_improvement} alongside the Adam optimizer \cite{2020_soydaner_comparison}.
 
\subsection{Impacts of parameter ranges on parameter estimation}\label{Impacts of parameter ranges on parameter estimation}
Due to the biological significance of the model parameters, it is crucial to set a range of each unknown parameter based on biological principles during estimation with \textit{DINNs}. Therefore, the unknown parameters are restricted to specific ranges so that the model can learn accurately from the experimental data. For instance, if the actual value of a parameter is 1, a search range of 10\% would correspond to $(1-0.1, 1+0.1)$, \textit{i.e.}, $(0.9, 1.1)$. In this way, a baseline value for each unknown parameter is selected randomly, falling within its predefined search range.
The model is trained on a dataset of 100 points over one million iterations using a cyclic learning rate ranging from \(10^{-6}\) to \(10^{-3}\), following the approach in \cite{2017_smith_cyclical}.
Initially, we test with various parameter ranges to understand their impacts on the outcome of the model.  
In this section, three numerical experiments are performed as follows:
\begin{itemize}
    \item \textbf{Experiment-1:} Parameters are estimated with a $10\%$ search range.
    \item \textbf{Experiment-2:} Parameters are estimated with a $20\%$ search range.
    \item \textbf{Experiment-3:} Parameters are estimated with a $50\%$ search range.
\end{itemize}
In these experiments, we consider a relative root mean squared error loss metric \text{`MSE NN'} which is defined as a tuple of relative root mean squared errors (RRMSE) computed for each dependent variable of the system \eqref{main model}. Specifically, for each dependent variable \( Z \in \{X, Y, D, V\} \), the relative root mean squared error is given by
\[
\text{RRMSE}_Z =\sqrt \frac{ \sum_{i=1}^{N} \left( \hat{Z}(t_i) - Z(t_i) \right)^2}{ \sum_{i=1}^{N} Z(t_i)^2}
\]
where \( \hat{Z}(t_i) \) denotes the prediction by \textit{DINNs} at time \( t_i \), and \( Z(t_i) \) is the corresponding observed data. The loss metric `MSE NN' is then defined as:
$
\text{MSE NN} = (\text{RRMSE}_X, \ \text{RRMSE}_Y, \ \text{RRMSE}_D, \ \text{RRMSE}_V).
$

\begin{figure}
\centering
\underline{\textbf{Prediction with 10$\%$ variation of the parameters}}
\includegraphics[height=4.5cm, width=17cm]{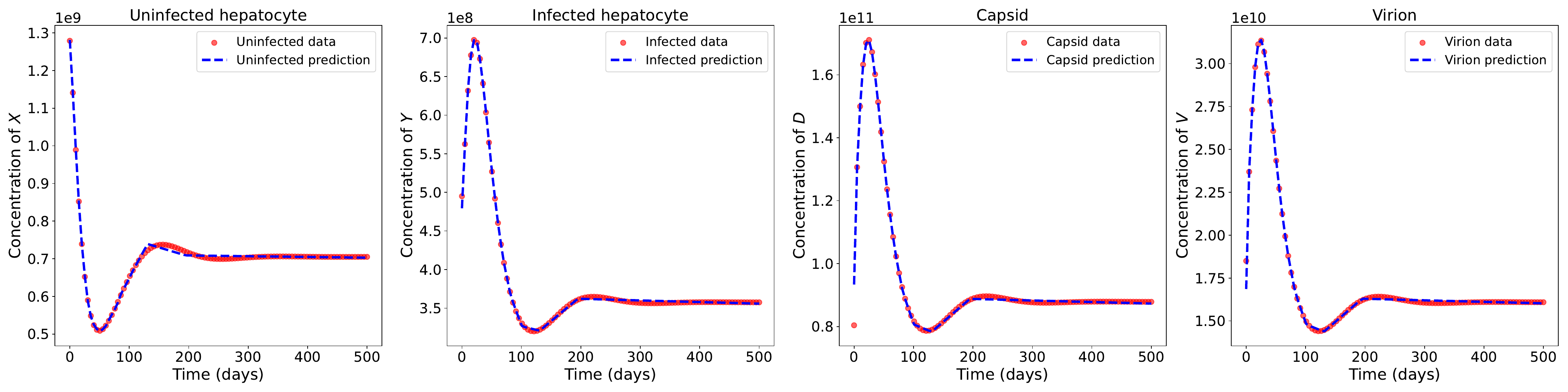}
\par\noindent\rule{\textwidth}{1pt}
\underline{\textbf{Prediction with 20$\%$ variation of the parameters}}
\includegraphics[height=4.5cm, width=17cm]{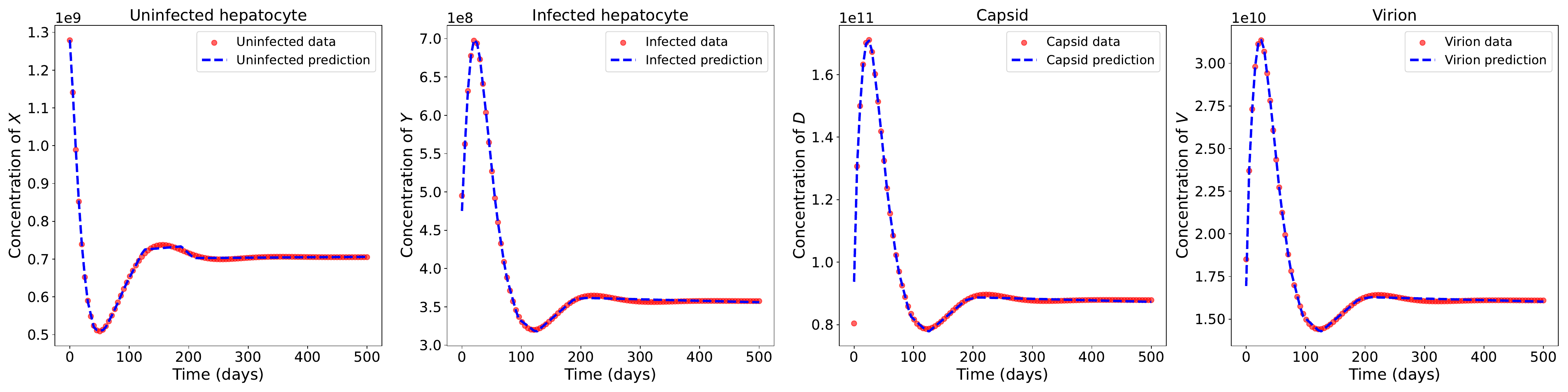}
\par\noindent\rule{\textwidth}{1pt}
\underline{\textbf{Prediction with 50$\%$ variation of the parameters}}
\includegraphics[height=4.5cm, width=17cm]{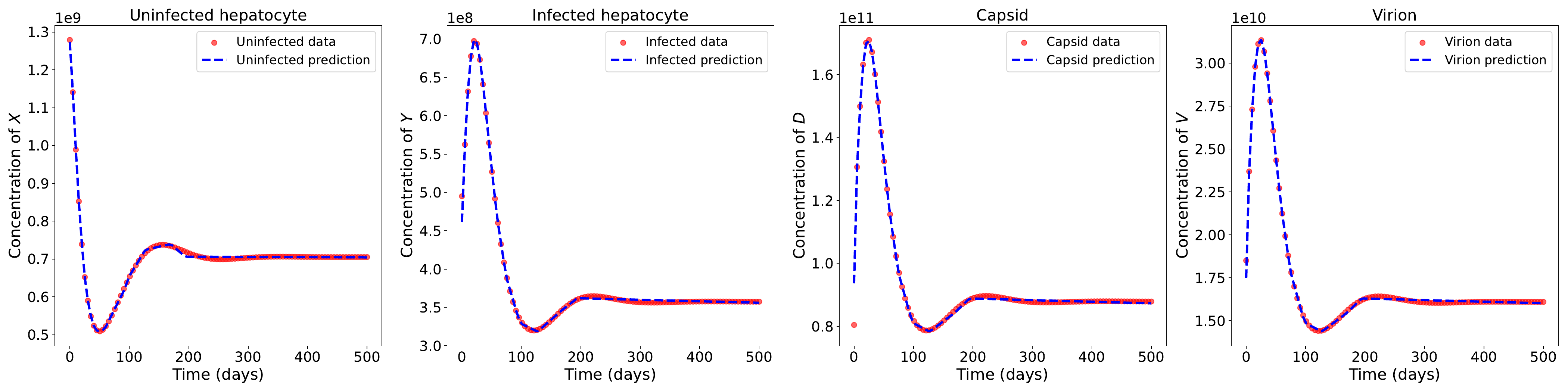}
\caption{\textit{DINNs} performance with varying parameter ranges.}
\label{fig: DINNs performance with varying parameter ranges}
\end{figure}

\noindent The first five columns of Table \ref{tab:optimal_parameters} present the model parameters, their actual values, the search ranges used by \textit{DINNs}, the values estimated by \textit{DINNs}, and the percentage relative errors between the actual and estimated values.
The rightmost column of Table \ref{tab:optimal_parameters} displays the `MSE' and `MSE NN' for each experiment.
\textit{DINNs} effectively identify the parameters sufficiently close to their actual values especially for the 10\% and 20\% search ranges as shown in Table \ref{tab:optimal_parameters}. When the search range is widened to 50\%, the estimated parameters exhibit greater variability (see Table \ref{tab:optimal_parameters}); however, the resulting solutions align well with the observed data (see Figure \ref{fig: DINNs performance with varying parameter ranges}).
Figure \ref{fig: DINNs performance with varying parameter ranges} displays the solutions of the system \eqref{main model} over time, obtained with 10\%, 20\% and 50\% variation in the search range. 
Although the system is trained well, small fluctuations in the `MSE NN' are observed across different experiments. Nevertheless, it is important to note that \textit{DINNs} are capable of learning the dynamics of the system \eqref{main model} in all three experiments, even when the learned parameter sets differ from each other. 
The possible reasons of this occurrence include several factors, such as relatively simpler diseases model, complex deep learning network. 
From these experiments, it can be concluded that \textit{DINNs} are capable of estimating different sets of parameters depending on the selected search range. Despite variations in the learned parameters, the model can capture the underlying dynamics of the system, producing solutions that reflect the patterns inherent in the observed data. This suggests that while parameter estimation using \textit{DINNs} with different search ranges may yield different sets of parameter values, \textit{DINNs} remain robust in learning meaningful and reliable representations of the system \eqref{main model}.

\begin{table}
\centering
\resizebox{\textwidth}{!}{%
\begin{tabular}{|c|c|c|c|c|c|c|}
\hline
\textbf{P} & \textbf{Actual Value} & \textbf{Range} & \textbf{PF} & \textbf{\% RE} & \multicolumn{2}{|c|}{\textbf{Errors}} \\ 
\hline
\multicolumn{7}{|c|}{\textbf{10\% Search Range}} \\ 
\hline
$\lambda$ & $2.6 \times 10^{7}$ & $(2.34 \times 10^{7}, 2.86 \times 10^{7})$ & $2.717 \times 10^{7}$ & 4.5 & \multirow{3}{*}{} & \multirow{3}{*}{} \\ \cmidrule(lr){1-5}
$\mu$ & 0.01 & (0.009, 0.011) & 0.0099 & 1 & & \\ \cmidrule(lr){1-5}
$k$ & $1.67 \times 10^{-12}$ & $(1.503 \times 10^{-12}, 1.837 \times 10^{-12})$ & $1.7525 \times 10^{-12}$ & 4.940 & & \\ 
\hline
$a$ & 150 & (135, 165) & 150.9204 & 0.614 & \textbf{MSE NN} & (0.0074, 0.0058, 0.0140, 0.0101) \\ 
\hline
$\beta$ & 0.87 & (0.793, 0.957) & 0.8608 & 1.057 & \textbf{MSE} & 0.0005 \\ 
\hline
$\delta$ & 0.053 & (0.0477, 0.0583) & 0.0547 & 3.208 & \multirow{4}{*}{} & \multirow{4}{*}{} \\ \cmidrule(lr){1-5}
$c$ & 3.8 & (3.42, 4.18) & 3.8204 & 0.537 & & \\ \cmidrule(lr){1-5}
$\alpha$ & 0.8 & (0.72, 0.88) & 0.8146 & 1.825 & & \\ \cmidrule(lr){1-5}
$\gamma$ & 0.6931 & (0.62379, 0.76241) & 0.7548 & 9.205 & & \\ 
\hline
\multicolumn{7}{|c|}{\textbf{20\% Search Range}}                                                                 \\ \hline
$\lambda$ & $2.6 \times 10^{7}$ & $(2.08 \times 10^{7}, 3.12 \times 10^{7})$ & $2.602 \times 10^{7}$ & 0.077 & \multirow{3}{*}{} & \multirow{3}{*}{} \\ \cmidrule(lr){1-5}
$\mu$ & 0.01 & (0.008, 0.012) & 0.0094 & 6 & & \\ \cmidrule(lr){1-5}
$k$ & $1.67 \times 10^{-12}$ & $(1.336 \times 10^{-12}, 2.005 \times 10^{-12})$ & $1.6500 \times 10^{-12}$ & 1.198 & & \\ 
\hline
$a$ & 150 & (120, 180) & 149.9617 & 0.026 & \textbf{MSE NN} & (0.0045, 0.0066, 0.0145, 0.0098) \\ 
\hline
$\beta$ & 0.87 & (0.696, 1.044) & 0.8514 & 3.287 & \textbf{MSE} & 0.0005  \\ 
\hline
$\delta$ & 0.053 & (0.0424, 0.0636) & 0.0513 & 3.208 & \multirow{4}{*}{} & \multirow{4}{*}{} \\ \cmidrule(lr){1-5}
$c$ & 3.8 & (3.04, 4.56) & 3.8322 & 0.847 & & \\ \cmidrule(lr){1-5}
$\alpha$ & 0.8 & (0.64, 0.96) & 0.8262 & 3.275 & & \\ \cmidrule(lr){1-5}
$\gamma$ & 0.6931 & (0.55448, 0.83172) & 0.8187 & 18.107 & & \\ 
\hline
\multicolumn{7}{|c|}{\textbf{50\% Search Range}} \\ 
\hline
$\lambda$ & $2.6 \times 10^{7}$ & $(1.3 \times 10^{7}, 3.9 \times 10^{7})$ & $2.292 \times 10^{7}$ & 11.846 & \multirow{3}{*}{} & \multirow{3}{*}{} \\ \cmidrule(lr){1-5}
$\mu$ & 0.01 & (0.005, 0.015) & 0.0078 & 22 & & \\ \cmidrule(lr){1-5}
$k$ & $1.67 \times 10^{-12}$ & $(0.835 \times 10^{-12}, 2.505 \times 10^{-12})$ & $1.4709 \times 10^{-12}$ & 11.922 & & \\ 
\hline
$a$ & 150 & (75, 225) & 143.9348 & 4.043 & \textbf{MSE NN} & (0.0047, 0.0095, 0.0143, 0.0072) \\ 
\hline
$\beta$ & 0.87 & (0.435, 1.305) & 0.8508 & 2.207 & \textbf{MSE} & 0.0005 \\ 
\hline
$\delta$ & 0.053 & (0.0265, 0.0795) & 0.0475 & 10.377 & \multirow{4}{*}{} & \multirow{4}{*}{} \\ \cmidrule(lr){1-5}
$c$ & 3.8 & (1.9, 5.7) & 3.8669 & 1.761 & & \\ \cmidrule(lr){1-5}
$\alpha$ & 0.8 & (0.4, 1.2) & 0.8338 & 4.225 & & \\ \cmidrule(lr){1-5}
$\gamma$ & 0.6931 & (0.34665, 1.03965) & 1.0113 & 45.909 & & \\ 
\hline
\end{tabular}%
}
\caption{Estimated values of the model parameters for different search ranges of parameters. Here, P: Parameters, PF: Parameter Found, and RE: Relative Error.}
\label{tab:optimal_parameters}
\end{table}

\subsection{Impacts of noise in experimental data on parameter estimation}
\label{Impacts of noise on parameter estimation}
\begin{figure}[ht!]
  \centering
  \begin{minipage}{0.48\textwidth}
    \centering
    \includegraphics[width=\linewidth, height=5.7cm]{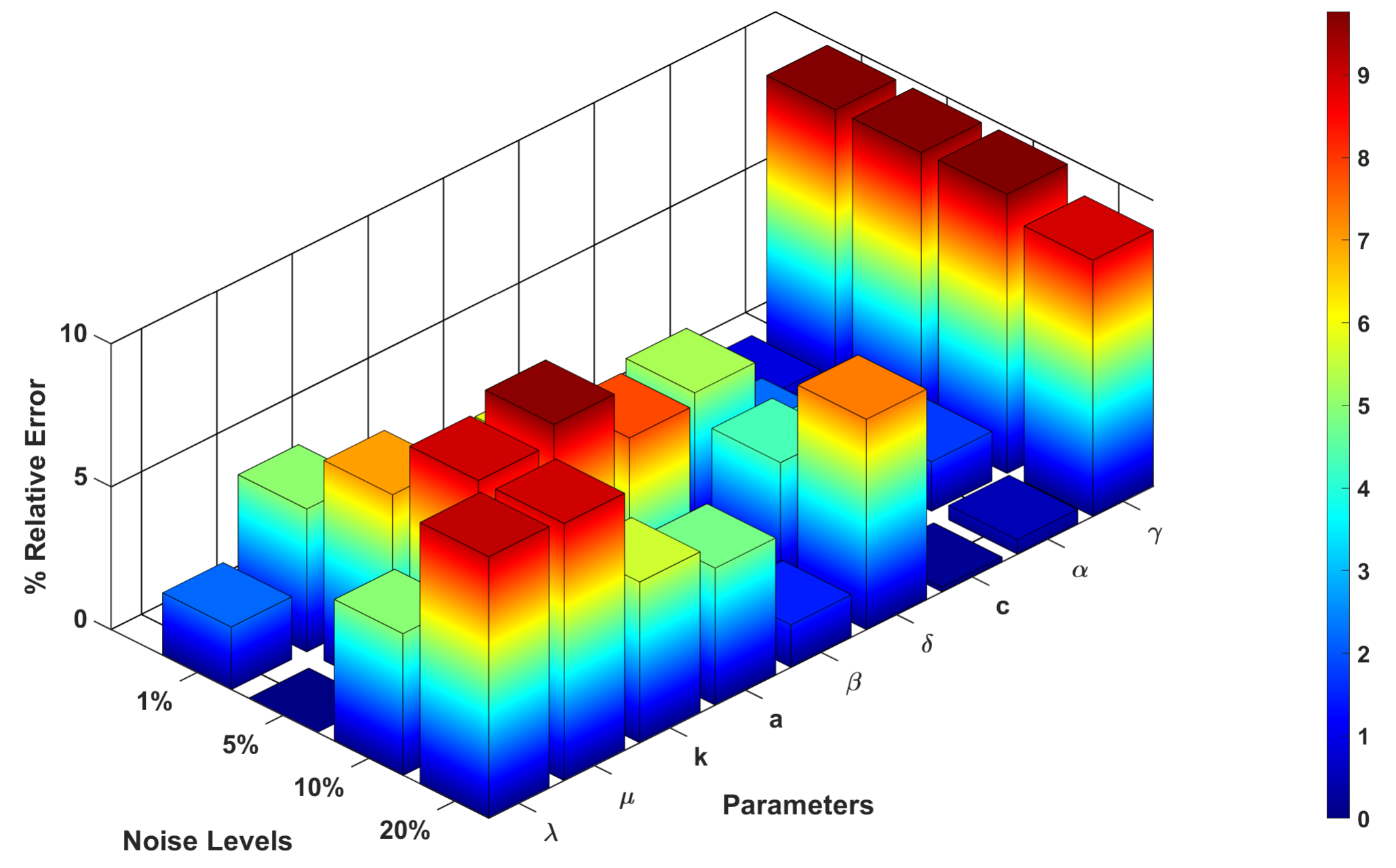}
    \caption{3D bar plot of the $\%$RE in the estimated parameters for different noise levels.}
    \label{3D bar plot}
  \end{minipage}%
  \hfill
  \begin{minipage}{0.48\textwidth}
    \centering
    \includegraphics[width=\linewidth, height=5.7cm]{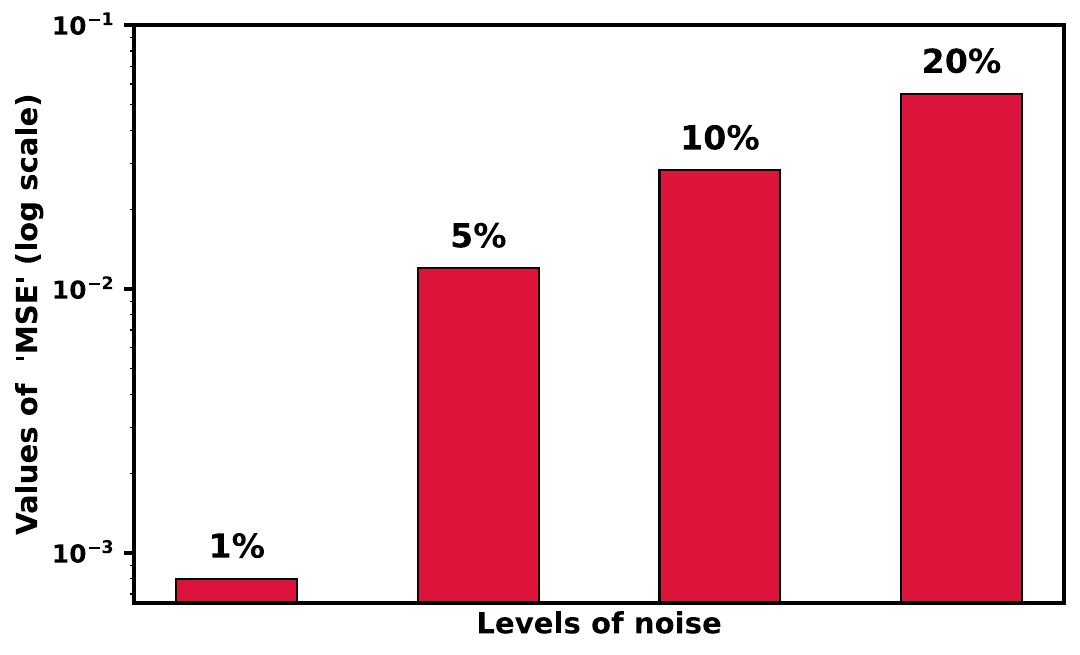}
    \caption{Bar plot of `MSE' for different levels of noise.}
    \label{fig:Bar plot of `MSE' for different levels of noise}
    \end{minipage}%
\end{figure}

\begin{figure}
\centering
\underline{\textbf{Prediction with 1\% gaussian noise}}
\includegraphics[height=4.5cm, width=17cm]{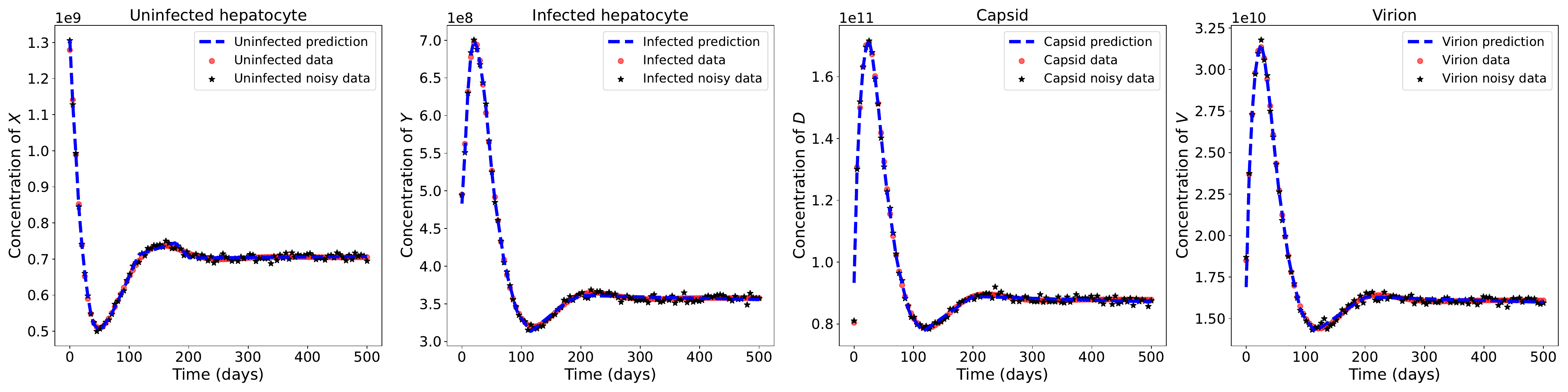}
\par\noindent\rule{\textwidth}{1pt}
\underline{\textbf{Prediction with 5\% gaussian noise}}
\includegraphics[height=4.5cm, width=17cm]{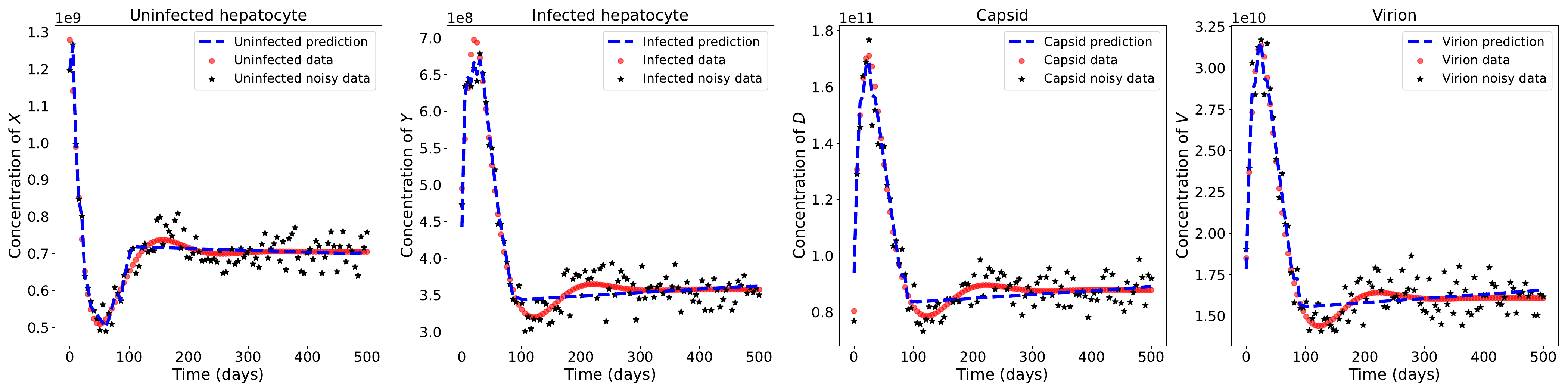}
\par\noindent\rule{\textwidth}{1pt}
\underline{\textbf{Prediction with 10\% gaussian noise}}
\includegraphics[height=4.5cm, width=17cm]{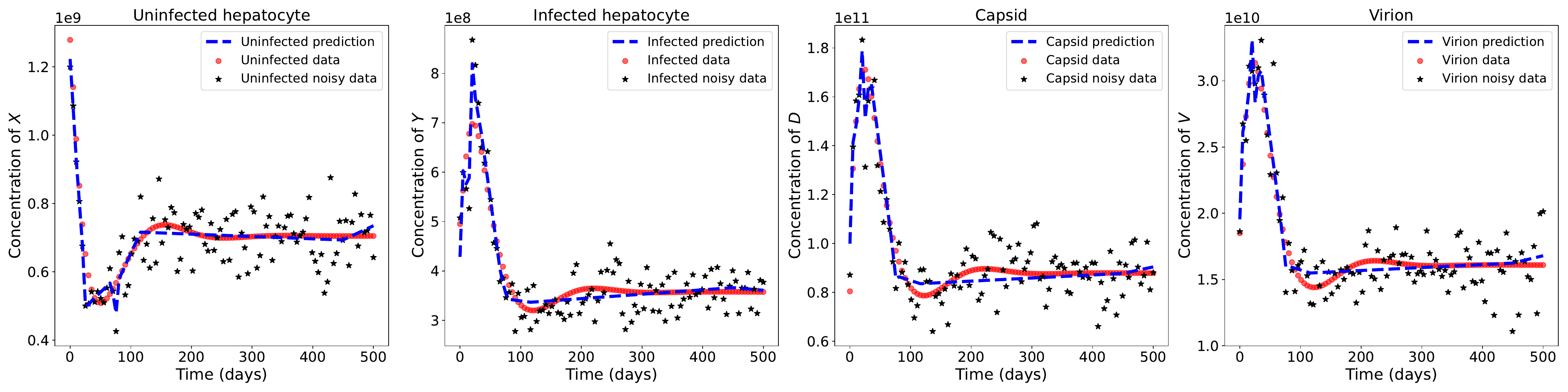}
\par\noindent\rule{\textwidth}{1pt}
\underline{\textbf{Prediction with 20\% gaussian noise}}
\includegraphics[height=4.5cm, width=17cm]{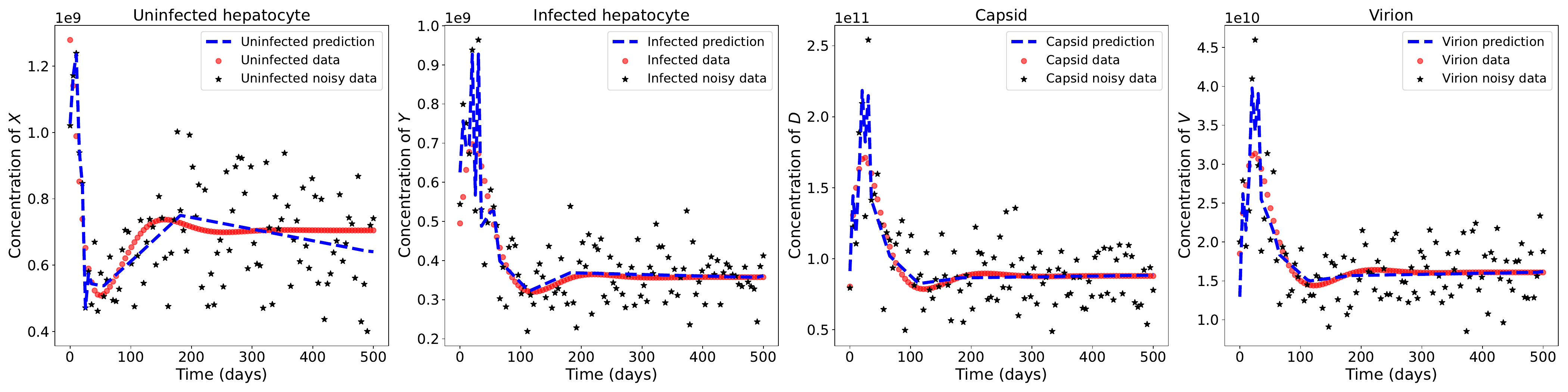}
\caption{\textit{DINNs} performance with varying uncorrelated gaussian noise.}
\label{fig: DINNs performance with varying uncorrelated gaussian noise}
\end{figure}

In case of biological phenomena, experimental data often contain inherent noise which can come from various sources such as measurement errors or experimental variability. 
This noise introduces uncertainties that can significantly affect the accuracy of parameter estimation. It is challenging to build a reliable model that can precisely capture the actual infection progression from a noisy dataset.
Noisy observations can obscure underlying dynamics and introduce uncertainty into the learning process, making it challenging for neural networks to accurately predict the temporal evolution of the dependent variables of a system. In order to confront these challenges and prove the reliability of \textit{DINNs}, different levels of uncorrelated gaussian noise (e.g., $1\%$, $5\%$, $10\%$, and $20\%$) are introduced into the data, and experiments are performed.
During these experiments, a 10\% variation in parameters is maintained, along with the same number of data points and learning rate as discussed in the previous Subsection \ref{Impacts of parameter ranges on parameter estimation}. These experiments reveal that \textit{DINNs} produce reliable outcomes under substantial noise levels up to 20\% and achieve a maximum RRMSE of $0.1228$ in learning the dynamics of system \eqref{main model}. However, the high level of noise complicates the precise learning of the dynamics, suggesting that additional training  is required to stabilize these estimates. The predictions made by \textit{DINNs} under 1\%, 5\%, 10\% and 20\% noise are visually represented in Figure \ref{fig: DINNs performance with varying uncorrelated gaussian noise} and  the optimal estimated values of the parameters are listed in Table \ref{tab:Noisy_data}. The $\%$RE in the estimated parameters and the variations in `MSE' across different levels of noise are depicted in Figures \ref{3D bar plot} and \ref{fig:Bar plot of `MSE' for different levels of noise}, respectively. The experiments in this section suggest that an increase in noise in the dataset decreases the ability of \textit{DINNs} to learn the system dynamics, leading to significant deviation in the values of estimated parameters from their true values.
\begin{table}
\centering
\begin{tabular}{|c|c|c|c|c|c|}
\hline
\textbf{P} & \textbf{Actual Value} & \textbf{PF} & \textbf{\% RE} & \multicolumn{2}{c|}{\textbf{Errors}} \\ \hline
\multicolumn{6}{|c|}{\textbf{1\% Uncorrelated Gaussian Noise}} \\ \hline
$\lambda$ & $2.6 \times 10^{7}$ & $2.657 \times 10^{7}$ & 2.192 & \multirow{3}{*}{} & \multirow{3}{*}{} \\ \cmidrule(lr){1-4}
$\mu$ & 0.01 & 0.0095 & 5  & & \\ \cmidrule(lr){1-4}
$k$ & $1.67 \times 10^{-12}$ & $1.702 \times 10^{-12}$ & 1.961 & &   \\ \hline
$a$ & 150 & 147.548 & 1.635 & \textbf{MSE NN} & (0.0077, 0.0006, 0.0143, 0.0104) \\ \hline
$\beta$ & 0.87 & 0.8608 & 1.057 &\textbf{MSE} & 0.0008 \\ \hline
$\delta$ & 0.053 & 0.0532 & 0.377 & \multirow{4}{*}{} & \multirow{4}{*}{} \\ \cmidrule(lr){1-4}
$c$ & 3.8 & 3.7870 & 0.342  & & \\ \cmidrule(lr){1-4}
$\alpha$ & 0.8 & 0.8071 & 0.888  & & \\ \cmidrule(lr){1-4}
$\gamma$ & 0.6931 & 0.7605 & 9.724 & &  \\ \hline
\multicolumn{6}{|c|}{\textbf{5\% Uncorrelated Gaussian Noise}} \\ \hline
$\lambda$ & $2.6 \times 10^{7}$ & $2.600 \times 10^{7}$ & 0 & \multirow{3}{*}{} & \multirow{3}{*}{} \\ \cmidrule(lr){1-4}
$\mu$ & 0.01 & 0.0093 & 7 & & \\ \cmidrule(lr){1-4}
$k$ & $1.67 \times 10^{-12}$ & $1.604 \times 10^{-12}$ & 3.952 & & \\ \hline
$a$ & 150 & 144.112 & 5.888 & \textbf{MSE NN} & (0.0228, 0.0351, 0.0311, 0.0295) \\ \hline
$\beta$ & 0.87 & 0.8633 & 0.770 &\textbf{MSE} & 0.0120 \\ \hline
$\delta$ & 0.053 & 0.0502 & 5.283 & \multirow{4}{*}{} & \multirow{4}{*}{} \\ \cmidrule(lr){1-4}
$c$ & 3.8 & 3.7155 & 2.224 & & \\ \cmidrule(lr){1-4}
$\alpha$ & 0.8 & 0.8016 & 0.200 & & \\ \cmidrule(lr){1-4}
$\gamma$ & 0.6931 & 0.7605 & 9.724 & & \\ \hline
\multicolumn{6}{|c|}{\textbf{10\% Uncorrelated Gaussian Noise}} \\ \hline
$\lambda$ & $2.6 \times 10^{7}$ & $2.471 \times 10^{7}$ & 4.962 & \multirow{3}{*}{} & \multirow{3}{*}{} \\ \cmidrule(lr){1-4}
$\mu$ & 0.01 & 0.0109 & 9 & & \\ \cmidrule(lr){1-4}
$k$ & $1.67 \times 10^{-12}$ & $1.831 \times 10^{-12}$ & 9.640 & & \\ \hline
$a$ & 150 & 138.228 & 7.848 & \textbf{MSE NN} & (0.0378, 0.0575, 0.0446, 0.0401) \\ \hline
$\beta$ & 0.87 & 0.8638 & 0.731 &\textbf{MSE} & 0.0282 \\ \hline
$\delta$ & 0.053 & 0.0553 & 4.339 & \multirow{4}{*}{} & \multirow{4}{*}{} \\ \cmidrule(lr){1-4}
$c$ & 3.8 & 3.6624 & 3.621 & & \\ \cmidrule(lr){1-4}
$\alpha$ & 0.8 & 0.7863 & 1.731 & & \\ \cmidrule(lr){1-4}
$\gamma$ & 0.6931 & 0.7608 & 9.768 & & \\ \hline
\multicolumn{6}{|c|}{\textbf{20\% Uncorrelated Gaussian Noise}} \\ \hline
$\lambda$ & $2.6 \times 10^{7}$ & $2.838 \times 10^{7}$ & 9.154 & \multirow{3}{*}{} & \multirow{3}{*}{} \\ \cmidrule(lr){1-4}
$\mu$ & 0.01 & 0.0091 & 9 & & \\ \cmidrule(lr){1-4}
$k$ & $1.67 \times 10^{-12}$ & $1.764 \times 10^{-12}$ & 5.629 & & \\ \hline
$a$ & 150 & 142.808 & 4.794 & \textbf{MSE NN} & (0.0756, 0.1228, 0.0898, 0.0949) \\ \hline
$\beta$ & 0.87 & 0.8571 & 1.483 &\textbf{MSE} & 0.0550 \\ \hline
$\delta$ & 0.053 & 0.0491 & 7.358 & \multirow{4}{*}{} & \multirow{4}{*}{} \\ \cmidrule(lr){1-4}
$c$ & 3.8 & 3.7939 & 0.1605 & & \\ \cmidrule(lr){1-4}
$\alpha$ & 0.8 & 0.8040 & 0.500 & & \\ \cmidrule(lr){1-4}
$\gamma$ & 0.6931 & 0.7552 & 8.959 & & \\ \hline
\end{tabular}%

\caption{Estimated values of the model parameters for different noise levels. Here, P: Parameters, PF: Parameter Found, and RE: Relative Error.}
\label{tab:Noisy_data}
\end{table}

\subsection{Impacts of sample sizes on parameter estimation}
\begin{figure}[h!]
\centering
\underline{\textbf{\textit{DINNs} output with 10 data points}}
\includegraphics[height=4.5cm, width=17cm]{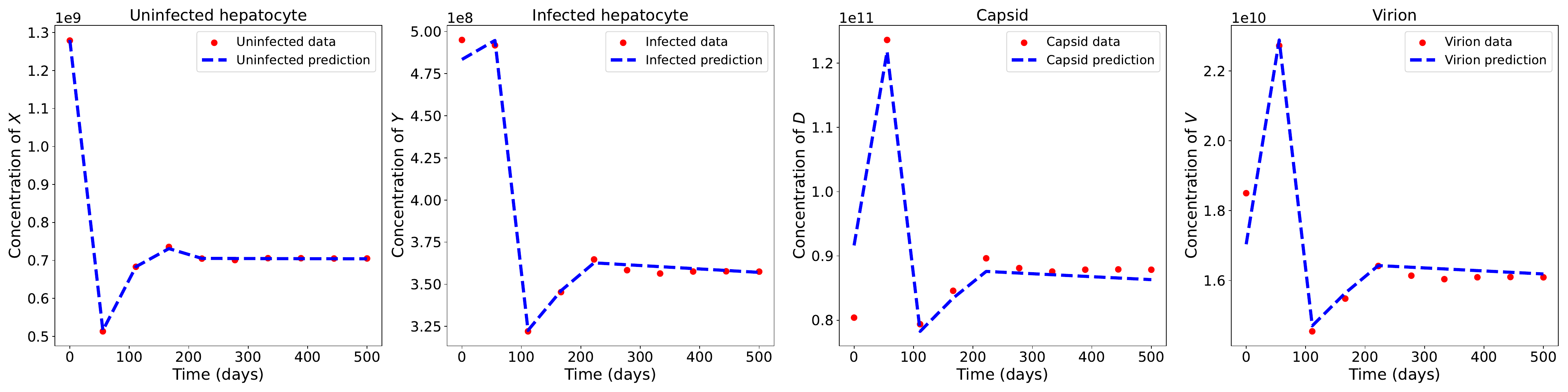}
\par\noindent\rule{\textwidth}{0.5pt}
\underline{\textbf{\textit{DINNs} output with 20 data points}}
\includegraphics[height=4.5cm, width=17cm]{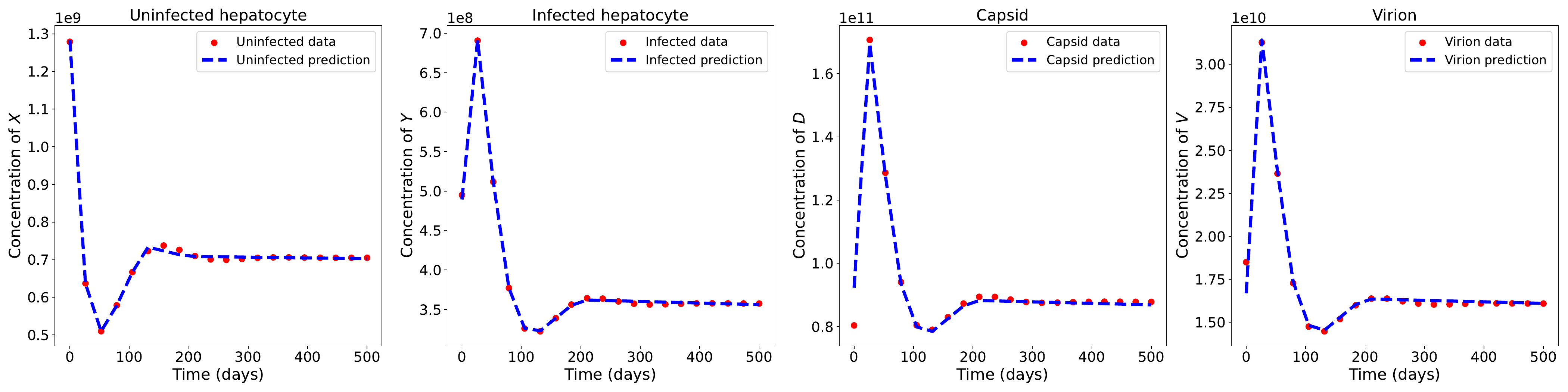}

\par\noindent\rule{\textwidth}{0.5pt}
\underline{\textbf{\textit{DINNs} output with 100 data points}}
\includegraphics[height=4.5cm, width=17cm]{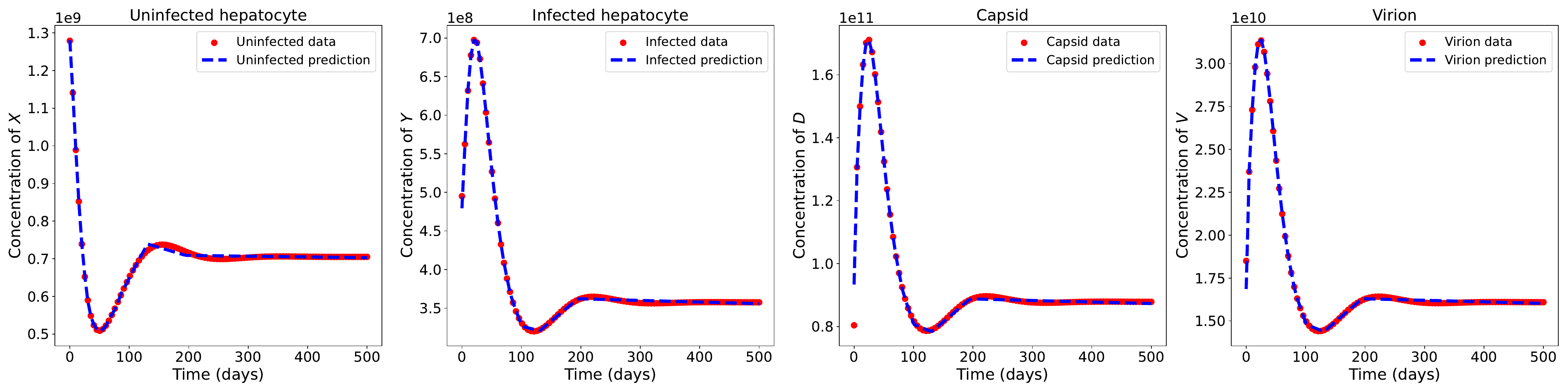}
\par\noindent\rule{\textwidth}{0.5pt}
\underline{\textbf{\textit{DINNs} output with 500 data points}}
\includegraphics[height=4.5cm, width=17cm]{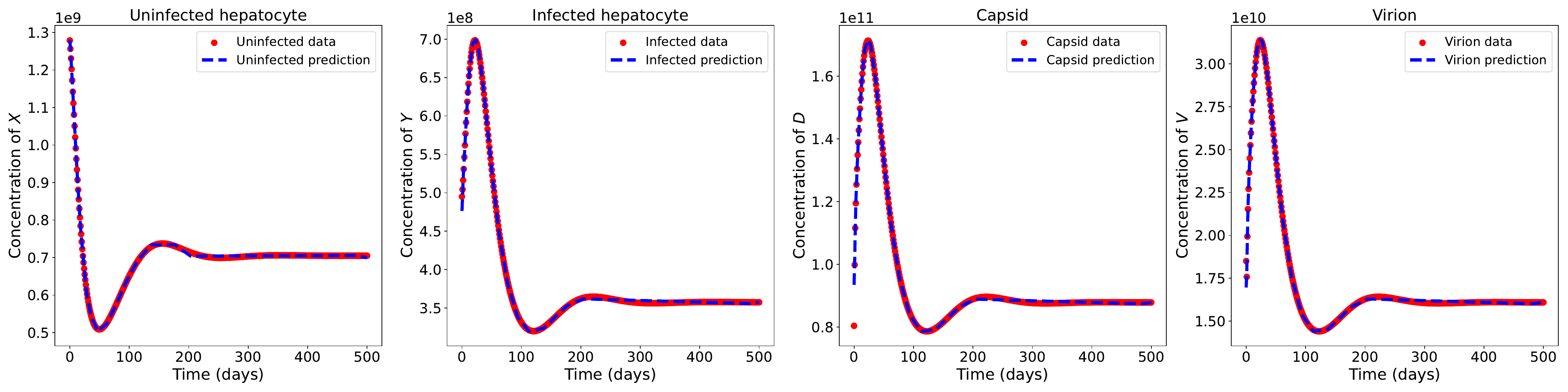}
\vspace{0.05cm}
\caption{\textit{DINNs} performance with varying data points.}
\label{fig: DINNs performance with varying data points}
\end{figure}

\begin{figure}[ht!]
  \centering
  \begin{minipage}{0.48\textwidth}
    \centering
    \includegraphics[width=\linewidth, height=5.7cm]{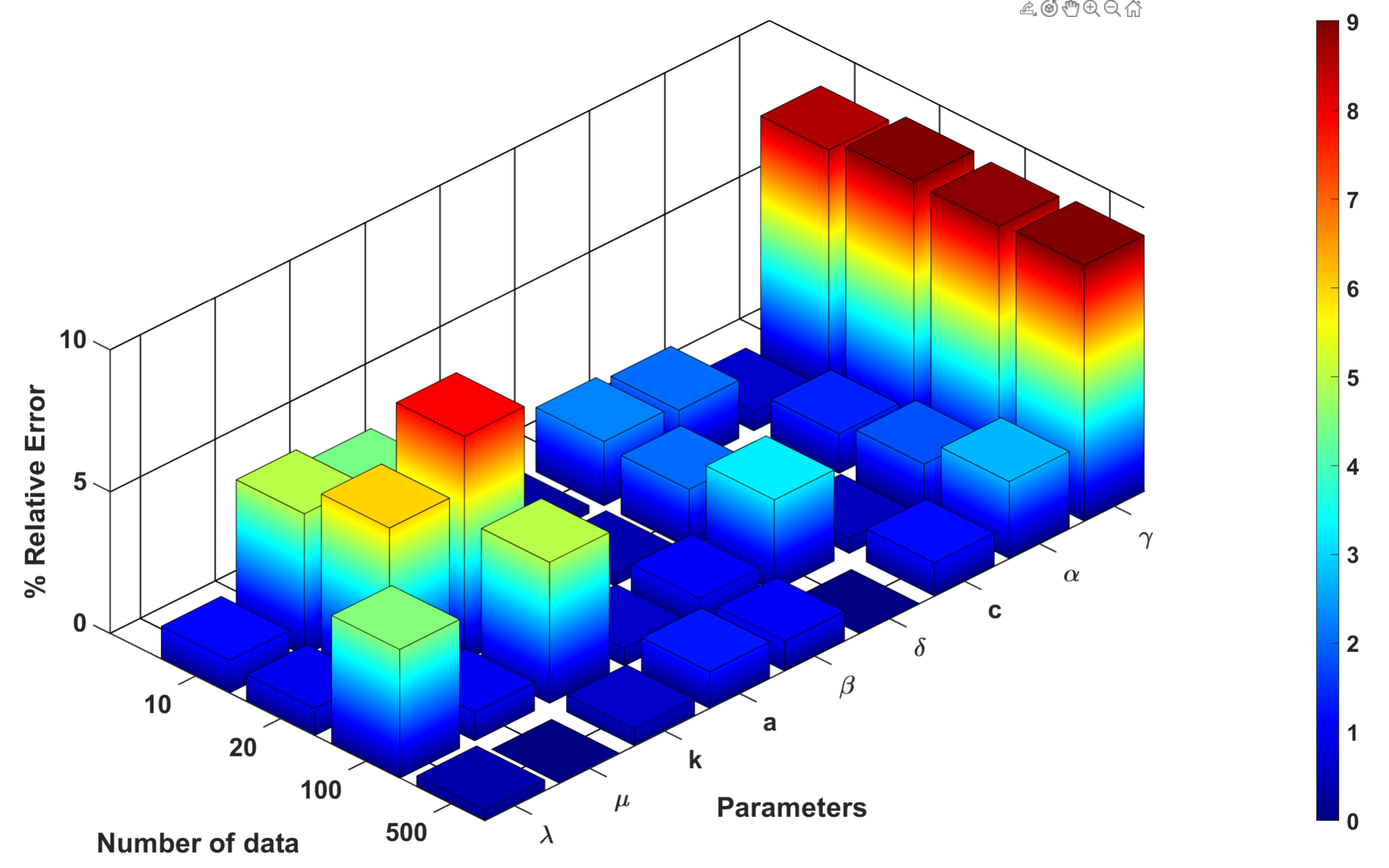}
    \caption{3D bar plot of the $\%$RE in the estimated parameters for different sample sizes.}
    \label{3D bar plot data}
  \end{minipage}%
  \hfill
  \begin{minipage}{0.48\textwidth}
    \centering
    \includegraphics[width=\linewidth, height=5.7cm]{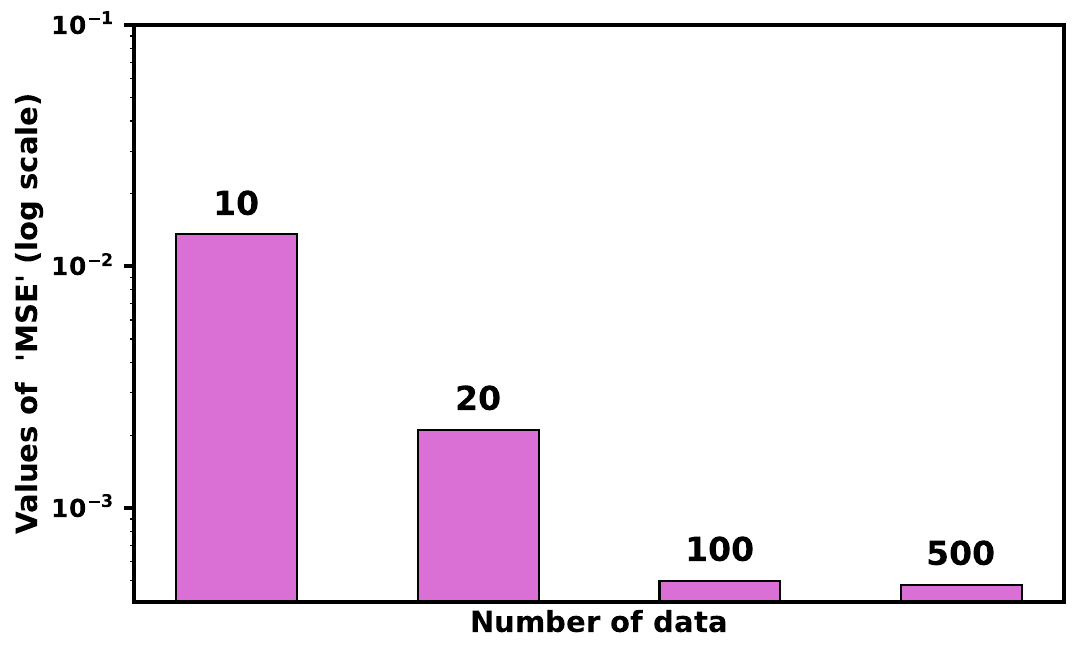}
    \caption{Bar plot of `MSE' for different sample sizes.}
    \label{fig:Bar plot of `MSE' for different sample sizes}
    \end{minipage}%
\end{figure}

For parameter estimation, adequate data are required to achieve reliable predictions. A sufficient amount of data helps the model to well-capture the underlying dynamics, while insufficient data leads to biased or inaccurate results \cite{2015_lecundeep}. This naturally leads to the question of how many data points are required for reliable training of \textit{DINNs}, and whether there exists a threshold beyond which additional data do not significantly enhance the model performance. In order to address this, we investigate the effects of data variability on `MSE' by training \textit{DINNs} with different sample sizes ranging from 10 to 500. Each dataset reflects different levels of information about the underlying phenomenon, allowing us to assess how the number of data impacts the performance of the model. When the model is trained with few data points (says 10), the model suffers to capture the dynamics of infected hepatocytes, capsids, and virions due to the scarcity of data (see Figure \ref{fig: DINNs performance with varying data points}). As the number of training data increases, the performance of the model improves, and at around 20 data points, the model becomes capable to learn both the dynamics of all the compartments and the associated parameters of the system \eqref{main model}. This finding suggests that for estimating the nine parameters of our ODE system \eqref{main model} and effectively capturing the dynamics of the dependent variables using \textit{DINNs}, around 20 data points are sufficient. The decision to consider of 20 data points is based on the result reported by Sontag \cite{2003_sontag_differential}.
The experiment using 100 data points yields an `MSE' of $5 \times 10^{-4}$, and no further significant improvement in `MSE' is observed when additional data are used for training. The values of estimated parameters, `MSE' and `MSE NN' from the experiments with 10, 20, 100 and 500 data points are shown in Table \ref{tab:Performance of DINNs for sample sizes}. For better visualization, 3D bar plot of the $\%$RE in the estimated parameters and 2D bar plot of `MSE' corresponding to the experiments with 10, 20, 100, and 500 data points are presented in Figures~\ref{3D bar plot data} and \ref{fig:Bar plot of `MSE' for different sample sizes}, respectively. 
This suggests that at around 100 data points, \textit{DINNs} becomes capable to capture well the dynamics of the system \eqref{main model}. However, this number may vary for other disease models, depending on their complexity and the number of unknown parameters. 
The experiments conducted in this section reinforce the notion that, in the task of parameter estimation, richer datasets generally lead to closer alignment with the true behavior of the system. However, beyond a certain threshold, addition of more data results in only marginal improvements in the performance of \textit{DINNs}.

\begin{table}[ht!]
\centering
\begin{tabular}{|c|c|c|c|c|c|}
\hline
\textbf{P} & \textbf{Actual Value} & \textbf{PF} & \textbf{\% RE} & \multicolumn{2}{c|}{\textbf{Errors}} \\ \hline
\multicolumn{6}{|c|}{\textbf{10 data points}} \\ \hline
$\lambda$ & $2.6 \times 10^{7}$ & $2.570 \times 10^{7}$ & 1.154 & \multirow{3}{*}{} & \multirow{3}{*}{} \\ \cmidrule(lr){1-4}
$\mu$ & 0.01 & 0.0105 & 5 & & \\ \cmidrule(lr){1-4}
$k$ & $1.67 \times 10^{-12}$ & $1.744 \times 10^{-12}$ & 4.431 & & \\ \hline
$a$ & 150 & 145.084 & 3.277 & \textbf{MSE NN} & (0.0029, 0.0109, 0.0416, 0.0290) \\ \hline
$\beta$ & 0.87 & 0.8676 & 0.276 & \textbf{MSE} & 0.0136 \\ \hline
$\delta$ & 0.053 & 0.0542 & 2.264 & \multirow{4}{*}{} & \multirow{4}{*}{} \\ \cmidrule(lr){1-4}
$c$ & 3.8 & 3.7223 & 2.045 & & \\ \cmidrule(lr){1-4}
$\alpha$ & 0.8 & 0.8053 & 0.663 & & \\ \cmidrule(lr){1-4}
$\gamma$ & 0.6931 & 0.7526 & 8.585 & & \\ \hline

\multicolumn{6}{|c|}{\textbf{20 data points}} \\ \hline
$\lambda$ & $2.6 \times 10^{7}$ & $2.625 \times 10^{7}$ & 0.962 & \multirow{3}{*}{} & \multirow{3}{*}{} \\ \cmidrule(lr){1-4}
$\mu$ & 0.01 & 0.0106 & 6 & & \\ \cmidrule(lr){1-4}
$k$ & $1.67 \times 10^{-12}$ & $1.802 \times 10^{-12}$ & 7.904 & & \\ \hline
$a$ & 150 & 149.115 & 0.590 & \textbf{MSE NN} & (0.0078, 0.0051, 0.0289, 0.0238) \\ \hline
$\beta$ & 0.87 & 0.8666 & 0.391 & \textbf{MSE} & 0.0021 \\ \hline
$\delta$ & 0.053 & 0.0519 & 2.075 & \multirow{4}{*}{} & \multirow{4}{*}{} \\ \cmidrule(lr){1-4}
$c$ & 3.8 & 3.7919 & 0.213 & & \\ \cmidrule(lr){1-4}
$\alpha$ & 0.8 & 0.8112 & 1.400 & & \\ \cmidrule(lr){1-4}
$\gamma$ & 0.6931 & 0.7554 & 8.989 & & \\ \hline

\multicolumn{6}{|c|}{\textbf{100 data points}} \\ \hline
$\lambda$ & $2.6 \times 10^{7}$ & $2.718 \times 10^{7}$ & 4.540 & \multirow{3}{*}{} & \multirow{3}{*}{} \\ \cmidrule(lr){1-4}
$\mu$ & 0.01 & 0.0099 & 1 & & \\ \cmidrule(lr){1-4}
$k$ & $1.67 \times 10^{-12}$ & $1.753 \times 10^{-12}$ & 4.970 & & \\ \hline
$a$ & 150 & 150.920 & 0.613 & \textbf{MSE NN} & (0.0073, 0.0058, 0.0140, 0.0101) \\ \hline
$\beta$ & 0.87 & 0.8608 & 1.057 & \textbf{MSE} & 0.0005 \\ \hline
$\delta$ & 0.053 & 0.0547 & 3.210 & \multirow{4}{*}{} & \multirow{4}{*}{} \\ \cmidrule(lr){1-4}
$c$ & 3.8 & 3.8204 & 0.537 & & \\ \cmidrule(lr){1-4}
$\alpha$ & 0.8 & 0.8146 & 1.825 & & \\ \cmidrule(lr){1-4}
$\gamma$ & 0.6931 & 0.7548 & 8.902 & & \\ \hline
\multicolumn{6}{|c|}{\textbf{500 data points}} \\ \hline
$\lambda$ & $2.6 \times 10^{7}$ & $2.610 \times 10^{7}$ & 0.385 & \multirow{3}{*}{} & \multirow{3}{*}{} \\ \cmidrule(lr){1-4}
$\mu$ & 0.01 & 0.0100 & 0 & & \\ \cmidrule(lr){1-4}
$k$ & $1.67 \times 10^{-12}$ & $1.680 \times 10^{-12}$ & 0.599 & & \\ \hline
$a$ & 150 & 151.9394 & 1.293 & \textbf{MSE NN} & (0.0037, 0.0046, 0.0072, 0.0056) \\ \hline
$\beta$ & 0.87 & 0.8609 & 1.046 & \textbf{MSE} & 0.00049 \\ \hline
$\delta$ & 0.053 & 0.0530 & 0 & \multirow{4}{*}{} & \multirow{4}{*}{} \\ \cmidrule(lr){1-4}
$c$ & 3.8 & 3.8407 & 1.070 & & \\ \cmidrule(lr){1-4}
$\alpha$ & 0.8 & 0.8174 & 2.715 & & \\ \cmidrule(lr){1-4}
$\gamma$ & 0.6931 & 0.7556 & 9.017 & & \\ \hline
\end{tabular}%

\caption{Estimated values of the model parameters for different sizes of data points. Here, P: Parameters, PF: Parameter Found, and RE: Relative Error.}
\label{tab:Performance of DINNs for sample sizes}
\end{table}

\subsection{Impacts of  network's architecture  on parameter estimation}
The architecture defines the structural design of a network, encompassing both the number of layers (depth) and the number of neurons in each layer (width). A proper architecture is of utmost important to determine how a network can accurately learn and reflect patterns inherent in data \cite{2019_shrestha_review,2016_sun_depth}. Deeper networks are more effective in learning intricate patterns from data than shallow ones.
However, careful tuning is required to prevent issues such as vanishing gradients or overfitting \cite{2024_xiong_eresnet}.
In some recent studies, it has been shown that expanding the width or depth of a neural network significantly affects the relative mean squared error between observed data and predicted values \cite{2017_guo_calibration,2020_lee_finite}. 
Based on this information, in this section, 
different combinations of network depth and width are considered to test and evaluate their impacts on the performance of the model. 
 The values of `MSE NN'  for the vector \( (X, Y, D, V) \) from these experiments are shown in Table  \ref{tab:Inﬂuence of Architecture Variation with $(X, Y, D, V)$ error from DINNs output}. These experiments demonstrate that increasing the width of the network consistently reduces the `MSE' in parameter estimation. In other words, when the number of layers is fixed, increasing the number of neurons in each layer leads to a steady improvement in model performance (see Figure \ref{`MSE' for different numbers of neurons per layer, with the number of layers fixed at 8.}).  
Furthermore, increasing the depth of the network initially results in a noticeable reduction in `MSE'. However, this change becomes insignificant beyond a certain threshold point, typically around eight layers. Beyond this, additional layers offer only marginal gains in the model's performance. This behavior suggests the presence of an optimal depth for the system \eqref{main model}. As depicted in Figure \ref{`MSE' with different numbers of layers, where each layer contains 64 neurons.}, the `MSE' decreases steadily with increasing depth until approximately eight layers, beyond which further increases in depth do not significantly improve the model's accuracy. These findings emphasize that balancing the depth and width is essential to reliably capture the dynamics and estimate the parameters of a system.

\begin{table}[ht!]
\centering
\resizebox{\textwidth}{!}{%
\begin{tabular}{|lllll|}
\hline
\multicolumn{1}{|l|}{\textbf{Layers}} & \multicolumn{1}{|c|}{\textbf{10 Neurons}} & \multicolumn{1}{|c|}{\textbf{20 Neurons}} & \multicolumn{1}{|c|}{\textbf{32 Neurons}} & \multicolumn{1}{|c|}{\textbf{64 Neurons}}  \\ \hline
\multicolumn{1}{|c|}{2} & \multicolumn{1}{l|}{$(0.0151, 0.0237, 0.0257, 0.0280)$} & \multicolumn{1}{l|}{$(0.0459, 0.0277, 0.0302, 0.0263)$} & \multicolumn{1}{l|}{$(0.0224, 0.0122, 0.0188, 0.0141)$} & \multicolumn{1}{l|}{$(0.0112, 0.0099, 0.0170, 0.0107)$} \\ \hline
\multicolumn{1}{|c|}{4} & \multicolumn{1}{l|}{$(0.0138, 0.0082, 0.0154, 0.0124)$} & \multicolumn{1}{l|}{$(0.0044, 0.0066, 0.0144, 0.0098)$} & \multicolumn{1}{l|}{$(0.0026, 0.0068, 0.0139, 0.0089)$} & \multicolumn{1}{l|}{$(0.0013, 0.0060, 0.0137, 0.0079)$} \\ \hline

\multicolumn{1}{|c|}{8} & \multicolumn{1}{l|}{$(0.0025, 0.0091, 0.0145, 0.0078)$} & \multicolumn{1}{l|}{$(0.0025, 0.0072, 0.0137, 0.0086)$} & \multicolumn{1}{l|}{$(0.0007, 0.0075, 0.0141, 0.0070)$} & \multicolumn{1}{l|}{$(0.0004, 0.0068, 0.0140, 0.0074)$} \\ \hline

\end{tabular}%
}

\caption{Impacts of network's architecture on the RRMSE of  $(X, Y, D, V)$ taken from \textit{DINNs} output.}
\label{tab:Inﬂuence of Architecture Variation with $(X, Y, D, V)$ error from DINNs output}
\end{table}

\begin{figure}[ht!]
  \centering
  \begin{minipage}{0.48\textwidth}
    \centering
    \includegraphics[width=\linewidth, height=6.5cm]{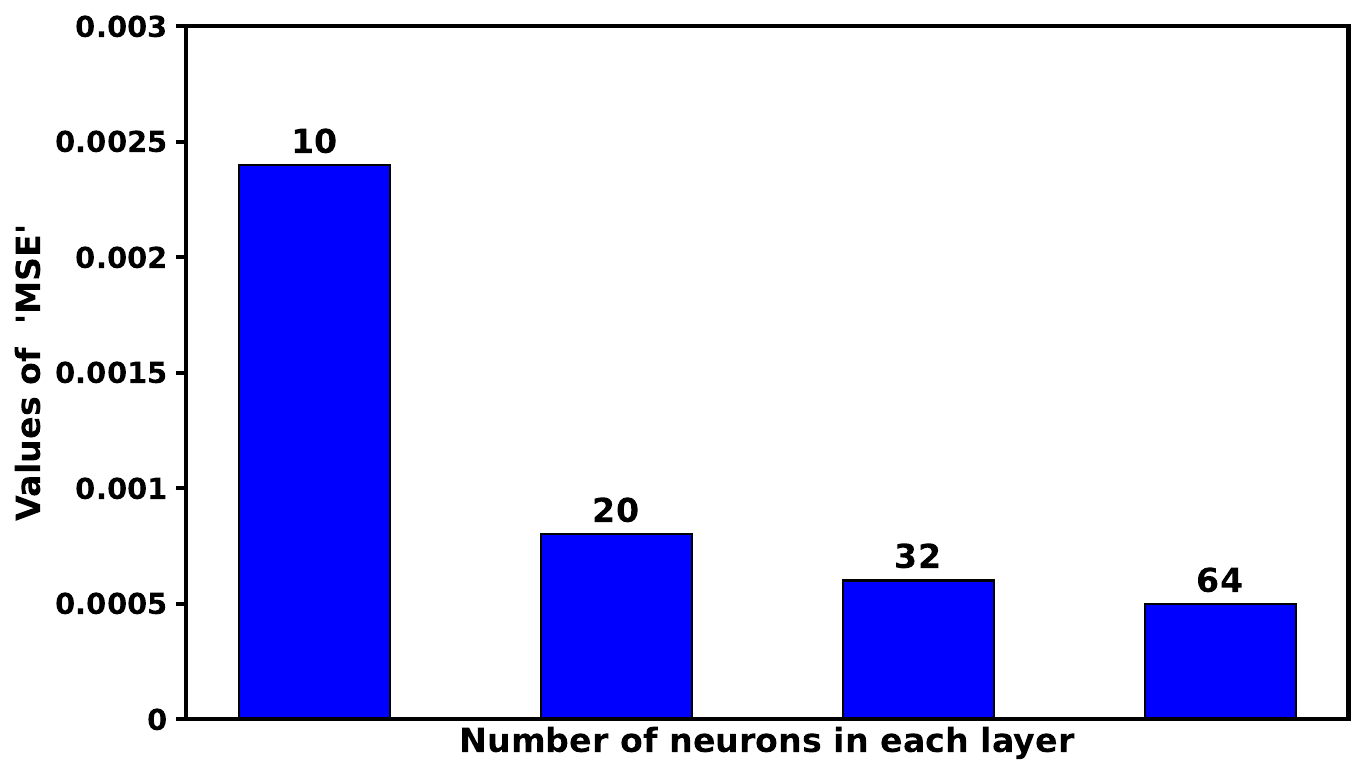}
    \caption{Bar plot of `MSE' for different numbers of neurons per layer, with the number of layers fixed at 8.}
    \label{`MSE' for different numbers of neurons per layer, with the number of layers fixed at 8.}
  \end{minipage}%
  \hfill
  \begin{minipage}{0.48\textwidth}
    \centering
    \includegraphics[width=\linewidth, height=6.5cm]{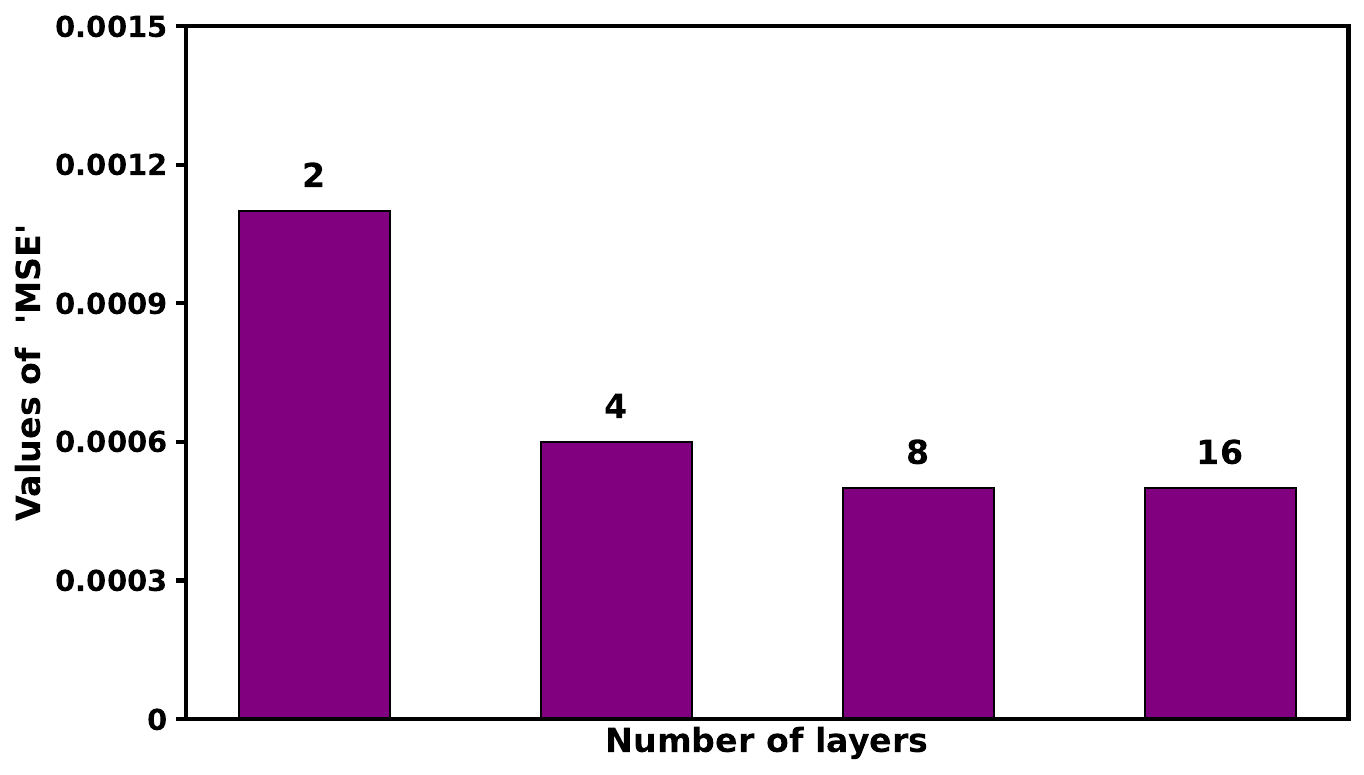}
    \caption{Bar plot of `MSE' with different numbers of layers, where each layer contains 64 neurons.}
    \label{`MSE' with different numbers of layers, where each layer contains 64 neurons.}
    \end{minipage}%
\end{figure}

\subsection{Impacts of learning rate on parameter estimation}

\begin{figure}[ht!]
  \centering
  \begin{minipage}{0.5\textwidth}
    \centering
    \includegraphics[width=\linewidth, height=6.5cm]{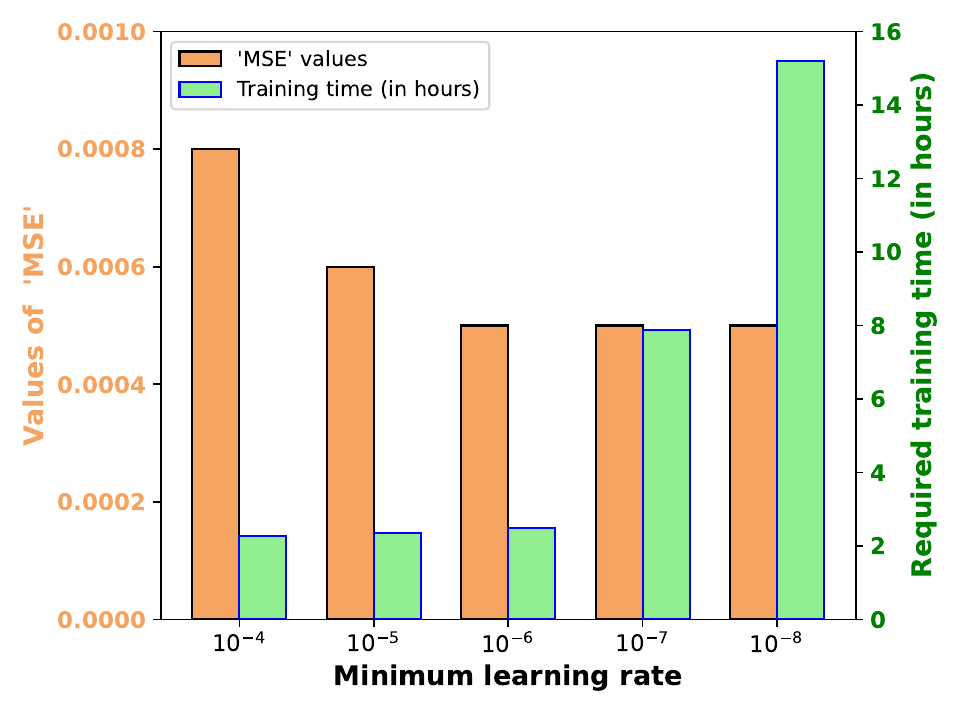}
    \caption{Bar plots of `MSE' and required training time for different values of minimum learning rate.}
    \label{fig:Effects of minimum learning rate on `MSE' and training time.}
  \end{minipage}%
  \hfill
  \begin{minipage}{0.48\textwidth}
    \centering
    \begin{tabular}{|C{2.5cm}|l|l|}
    \hline
    \textbf{ Minimum learning rate}    & \textbf{MSE} & \textbf{Training time} \\ \hline
    \textbf{$10^{-4}$}    & 0.0008 & 2 h 17 min \\ \hline
    \textbf{$10^{-5}$}    & 0.0006 & 2 h 22 min \\ \hline
    \textbf{$10^{-6}$}    & 0.0005 & 2 h 29 min \\ \hline
    \textbf{$10^{-7}$}    & 0.0005 & 7 h 53 min \\ \hline
     \textbf{$10^{-8}$}   & 0.0005 & 15 h 12 min \\ \hline
    \end{tabular}
    \captionof{table}{Impacts of minimum learning rate on `MSE' and training time.}
    \label{tab:lr_time}
  \end{minipage}
\end{figure}

The learning rate (LR) is a key hyperparameter for training deep neural networks. It controls how much the weights of a network need to be updated at each step. It directly influences the minimization of the loss function, thereby affecting both the convergence speed and the overall accuracy of the model.
A small LR makes only minor adjustments to the weights at each step. This ensures a stable and precise convergence of the model, but the training can be very slow. On the other hand, a large learning rate allows gradient descent to proceed more quickly, potentially leading to faster convergence. However, an excessively high LR will make the training algorithm diverge. Therefore, choosing an appropriate learning rate becomes a critical focus in deep learning. 
In some recent studies, it has been shown that starting with a high LR and slowly reducing it to a fixed value improves the accuracy in network's learning \cite{2021_nakamura_learning}.
A well-known example of such an LR schedule is PyTorch’s \text{CyclicLR} scheduler \cite{2024_shao_improvement}, which cyclically adjusts the learning rate based on a  predefined step size. In this schedule, one needs to specify the minimum and maximum limits, and the learning rate cyclically varies between these limits throughout the training. In order to assess the importance of LR in determining the required training time, five experiments are carried out using \textit{DINNs}. In these experiments, the maximum learning rate is set to $10^{-3}$ based on \cite{2020_gulde_deep} where minimum learning rates are chosen as \( 10^{-4}\), \( 10^{-5}\), \( 10^{-6}\), \( 10^{-7}\) and \( 10^{-8}\). Furthermore, the step size is fixed at $1000$. Initially, all experiments are conducted over one million iterations. The results indicate that this number of iterations is sufficient for the first three experiments to achieve convergence and estimate the model parameters within the predefined search range. Additionally, as the minimum learning rate decreases, a slight increase in training time is observed, accompanied by a small reduction in `MSE' for these three experiments. On the other hand, for the last two experiments, the model fails to learn the dynamics of \eqref{main model} and consequently is not able to estimate the parameters when trained for one million iterations, indicating that additional training is required in these cases. Therefore, to ensure convergence in the last two experiments, the model is trained further, requiring approximately 8 and 15 hours, respectively. However, this extended training yields no significant reduction in `MSE' compared to the experiment using a minimum learning rate of $10^{-6}$. The variations in `MSE' and the corresponding training times across different minimum learning rates are illustrated in Figure~\ref{fig:Effects of minimum learning rate on `MSE' and training time.} and summarized in Table~\ref{tab:lr_time}.
Overall, the results in this section highlight that while decreasing the minimum learning rate can slightly improve model performance, rapidly reducing it to an extremely small value significantly increases the training time.

\section{Applications of \textit{DINNs} to real-data}

\subsection{Data sources}

 In order to establish the  realism of viral dynamics and enhance the reliability of model predictions,  it is essential to validate the proposed model with the experimental data. Rather than relying on a single dataset, using multiple datasets offers a more robust and generalizable assessment of the model's performance. Biological systems often exhibit individual variability under diverse  experimental conditions. Therefore, in order to ensure the learned dynamics are not over-fitted, and reflect the underlying mechanisms governing HBV infection, it is important to validate the model across multiple datasets. To this purpose, experimental data of HBV DNA from nine young, healthy, HBV-seronegative chimpanzees are collected from the study of Asabe et al. \cite{2009_asabe_size} using a sophisticated software, Origin.  In their experiment, chimpanzees were infected with varying inoculation doses, ranging from $10^0$ to $10^{10}$. Asabe et al. \cite{2009_asabe_size} reported that out of the nine chimpanzees, six (labeled by ChA006, ChA007, Ch1622, ChA3A005, Ch1618 and ChA014) became acutely infected and  recovered from infection within a few weeks from the day of inoculation. On the other hand, chronic infection was developed in the remaining three chimpanzees (labeled by ChA2A007, Ch1603 and Ch1616). In this study, experimental data from both types of chimpanzees (acutely and chronically infected) are utilized to calibrate the proposed model. 
 
 \subsection{Methodology}
 Based on the structure and size of the experimental data, the entire data set for each chimpanzee is divided into two parts. Approximately, $(80-90)\%$ of the data is used to train the neural network,
 while the remaining $(10-20)\%$ is used to assess how well the model capture the actual HBV DNA levels beyond the training interval. The second column of Table \ref{tab:chimpanzee data} contains the initial inoculation doses administered to each chimpanzee, while the last three columns show the number of available experimental data, the number of data points used for training, and the number of data points kept for testing the performance of \textit{DINNs} beyond the training interval.

\begin{table}[h!]
\centering
\resizebox{\textwidth}{!}{%
\begin{tabular}{|l|C{3cm}|C{2cm}|C{2cm}|C{2cm}|}
\hline
\textbf{Chimpanzees} & \textbf{Initial inoculation dose (GE/ml)} & \textbf{Total available data} & \textbf{Data used for training} & \textbf{Data used for testing} \\
\hline
ChA006    & $10^{10}$ & 19 & 15 & 4 \\  \hline
ChA007    & $10^{7}$  & 11 & 9  & 2 \\  \hline
Ch1622    & $10^{4}$  & 14 & 11 & 3 \\  \hline
ChA3A005  & $10^{4}$  & 24 & 20 & 4 \\  \hline
Ch1618    & $10^{4}$  & 21 & 18 & 3 \\  \hline
ChA014    & $10^{1}$  & 27 & 22 & 5 \\  \hline
ChA2A007  & $10^{1}$  & 42 & 36 & 6 \\  \hline
Ch1603    & $10^{0}$  & 30 & 26 & 4 \\  \hline
Ch1616    & $10^{0}$  & 48 & 41 & 7 \\
\hline
\end{tabular}%
}
\caption{Summary of initial inoculation doses, number of available data points, number of data points used for training, and number of data points used for testing for each chimpanzee.}
\label{tab:chimpanzee data}
\end{table}
The network undergoes training for seven million iterations which is necessary to effectively learn the dynamics of HBV DNA-containing capsids. In these experiments, the learning rate used ranges from \(10^{-6}\) to \(10^{-3}\). Since experimental data are available only for HBV DNA-containing capsids, the data loss is defined here as:
\begin{equation}
\text{MSE}_{\text{Net\_HBV}} =  \frac{1}{N_u} \sum_{i=1}^{N_u} \left( D^{i} - \hat{D}^{i}\left(\hat{p}, \hat{\theta}\right)\right)^2.
\end{equation}
In this formulation, $D^i$ and $\hat{D}^{i}(\hat{p}, \hat{\theta})$ denote the observed and predicted concentrations of capsids at time point $t^i$ ($i = 1, 2, \ldots, N_u$), where $N_u$ is the the total number of available experimental data used for training. On the other hand, the residual loss term is taken same as defined in equation \eqref{MSE_NETF}. The overall MSE is then computed as the sum of data loss and residual loss, and it is minimized using the Adam optimizer \cite{2020_soydaner_comparison}.

\subsection{Experimental validation and forecast}
\begin{figure}[ht!]
\centering
\begin{subfigure}[b]{0.3\textwidth}
    \centering
    \includegraphics[height=5.3cm, width=5.7cm]{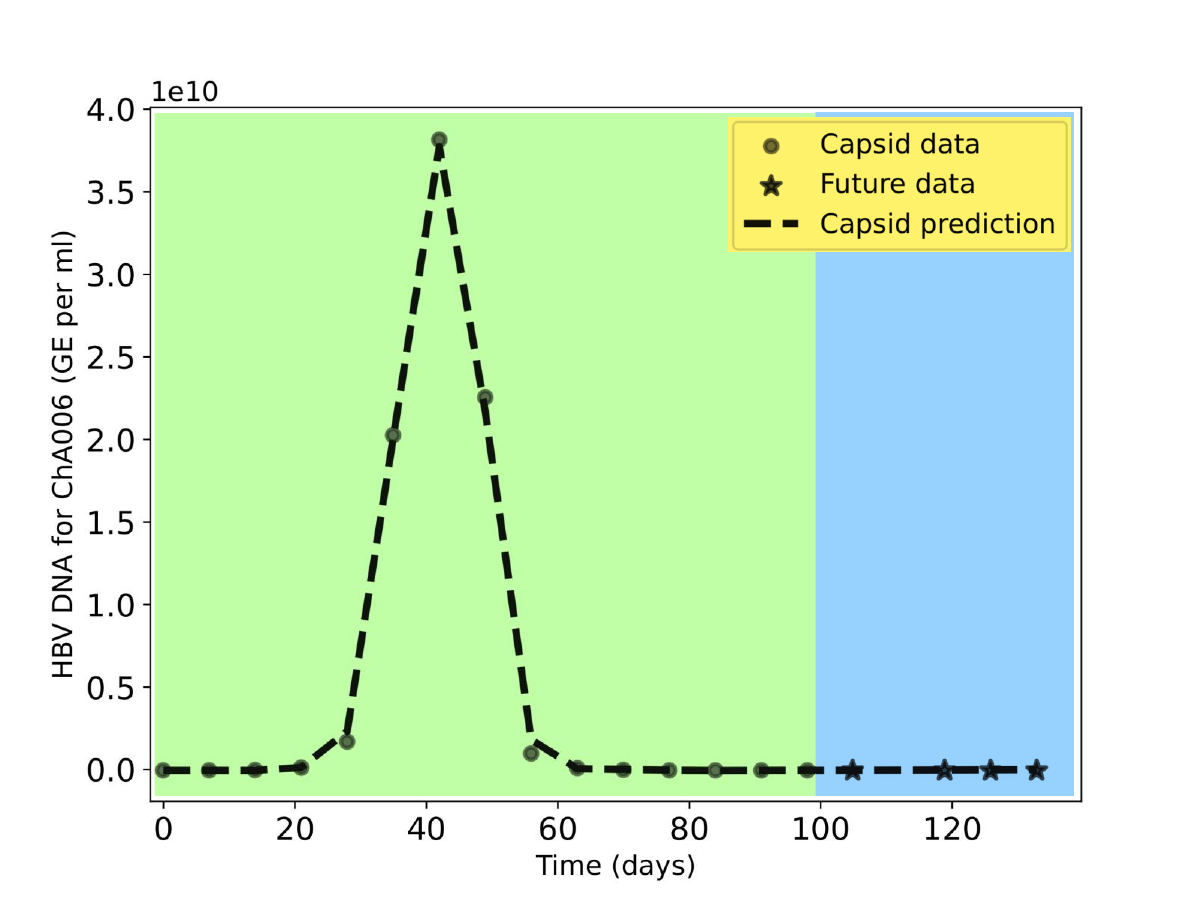}
    \caption{ChA006}
\end{subfigure}
\hfill
\begin{subfigure}[b]{0.3\textwidth}
    \centering
    \includegraphics[height=5.3cm, width=5.7cm]{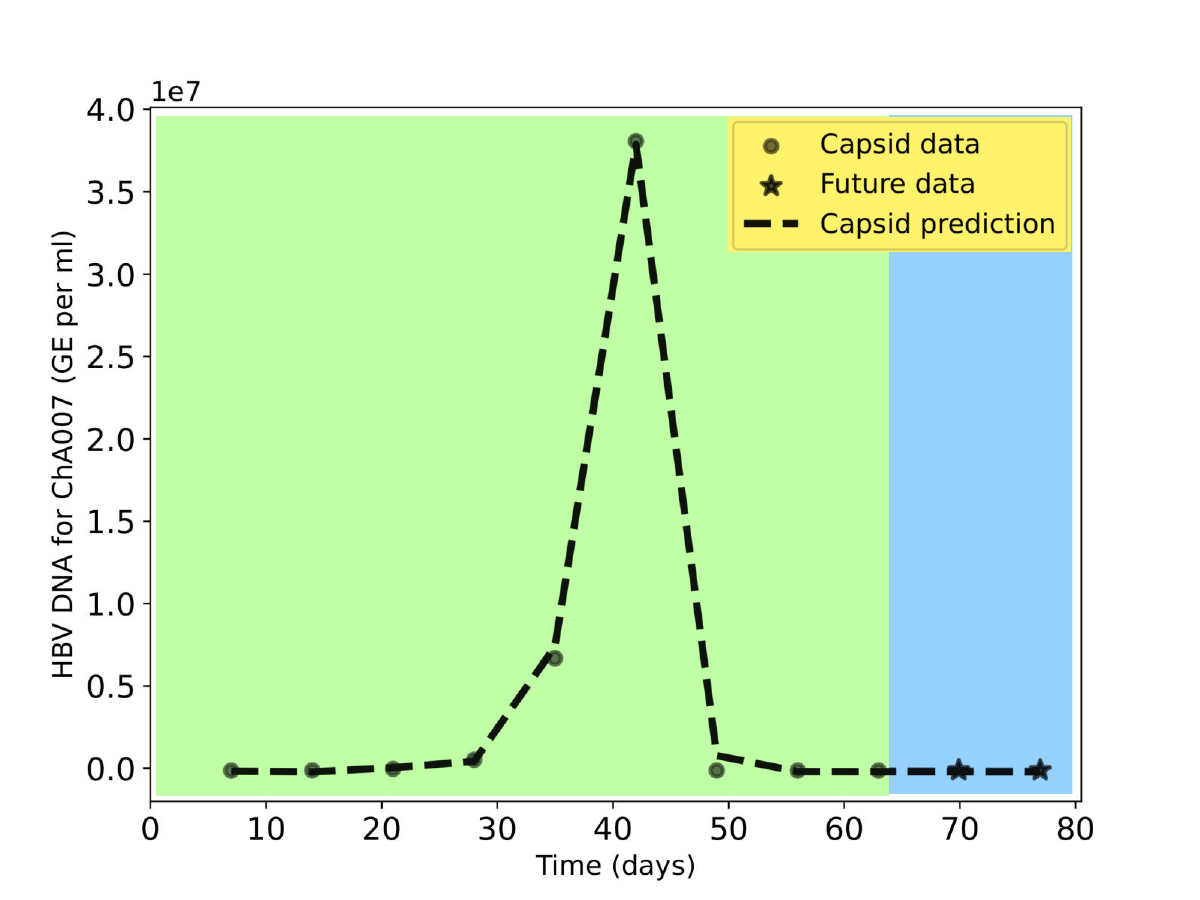}
    \caption{ChA007}
\end{subfigure}
\hfill
\begin{subfigure}[b]{0.3\textwidth}
    \centering
    \includegraphics[height=5.3cm, width=5.7cm]{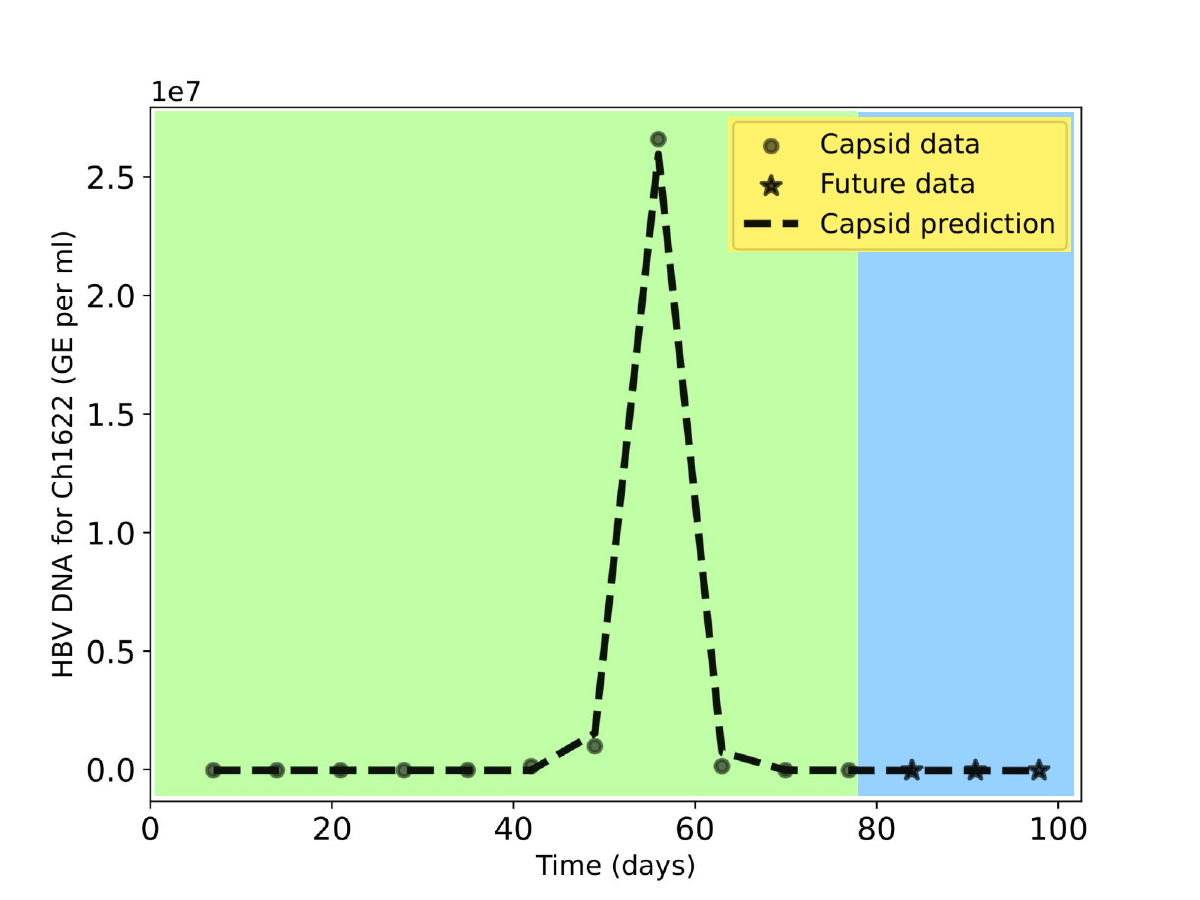}
    \caption{Ch1622}
\end{subfigure}
\hfill
\begin{subfigure}[b]{0.3\textwidth}
    \centering
    \includegraphics[height=5.3cm, width=5.7cm]{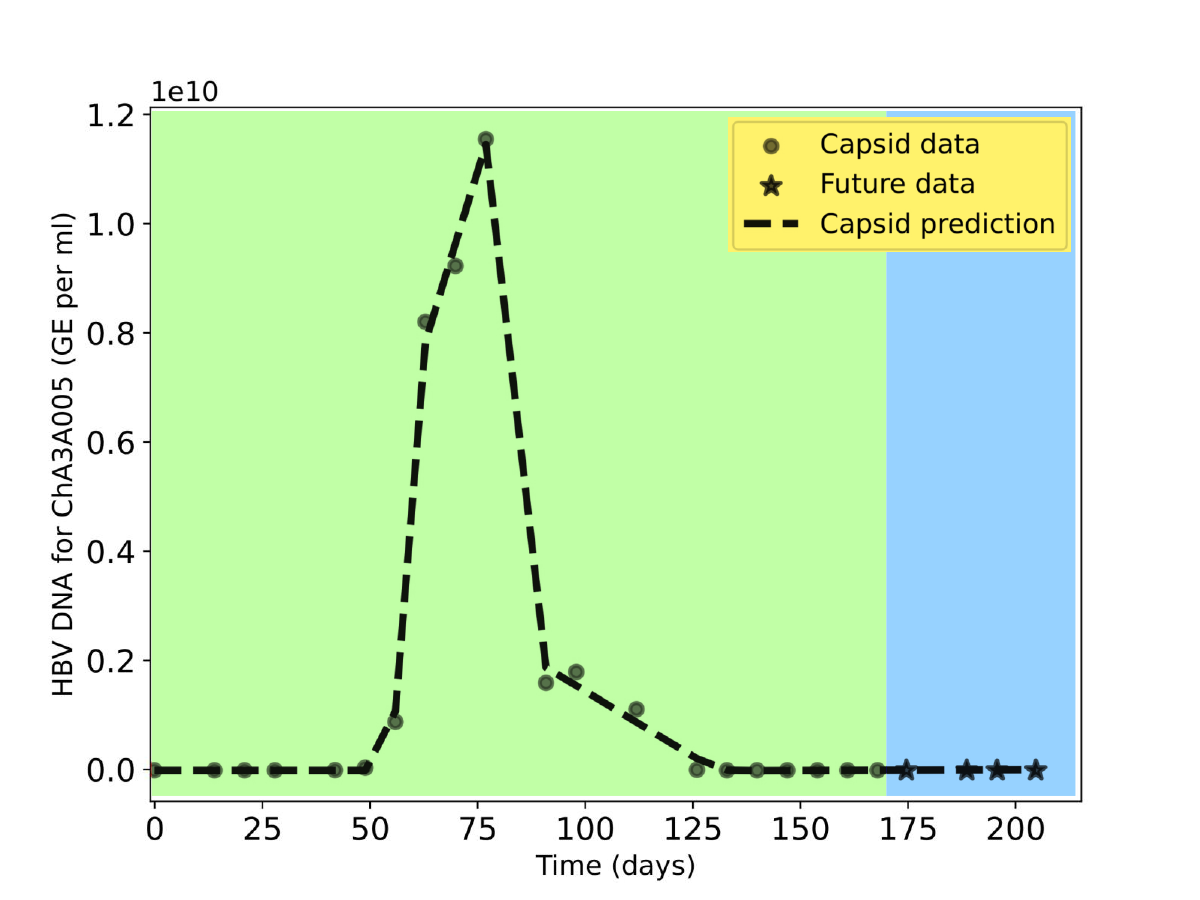}
    \caption{ChA3A005}
\end{subfigure}
\hfill
\begin{subfigure}[b]{0.3\textwidth}
    \centering
    \includegraphics[height=5.3cm, width=5.7cm]{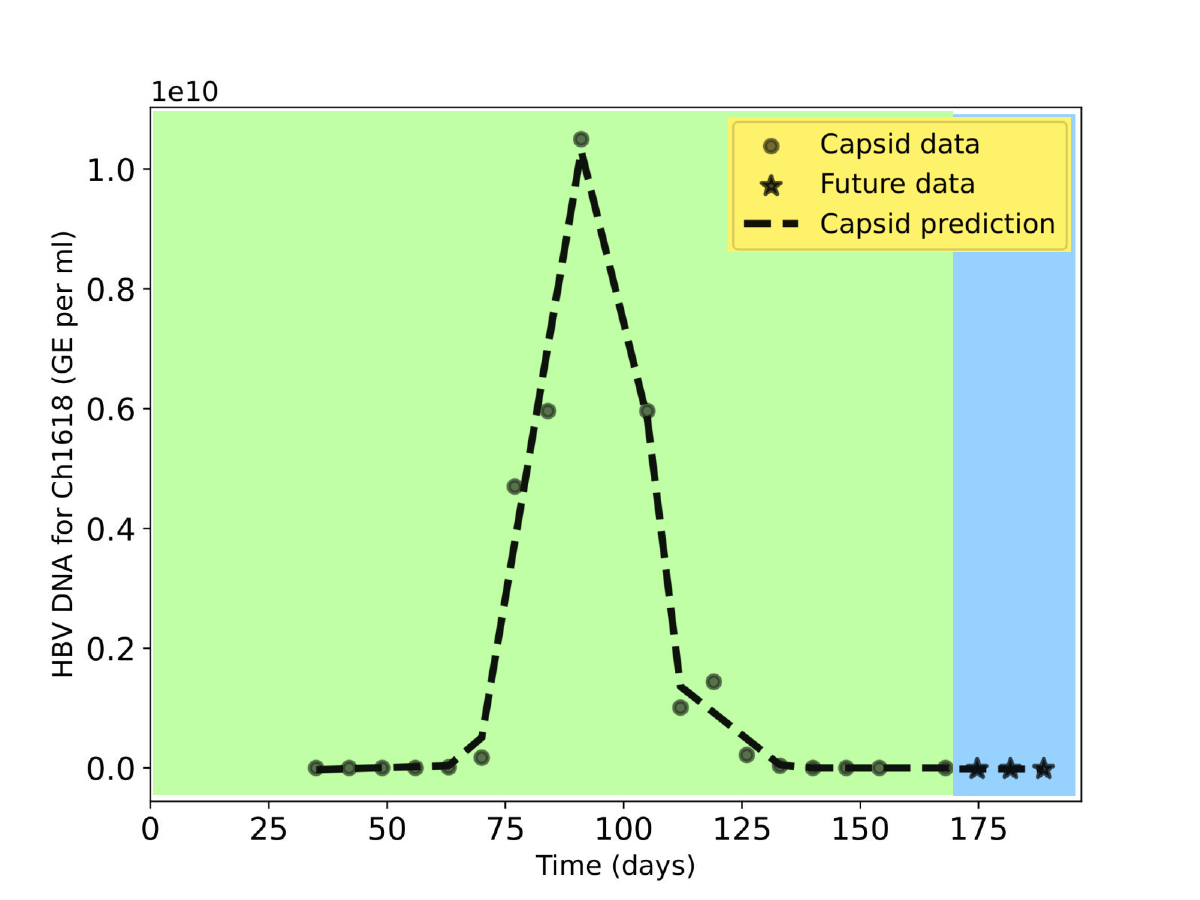}
    \caption{Ch1618}
\end{subfigure}
\hfill
\begin{subfigure}[b]{0.3\textwidth}
    \centering
    \includegraphics[height=5.3cm, width=5.7cm]{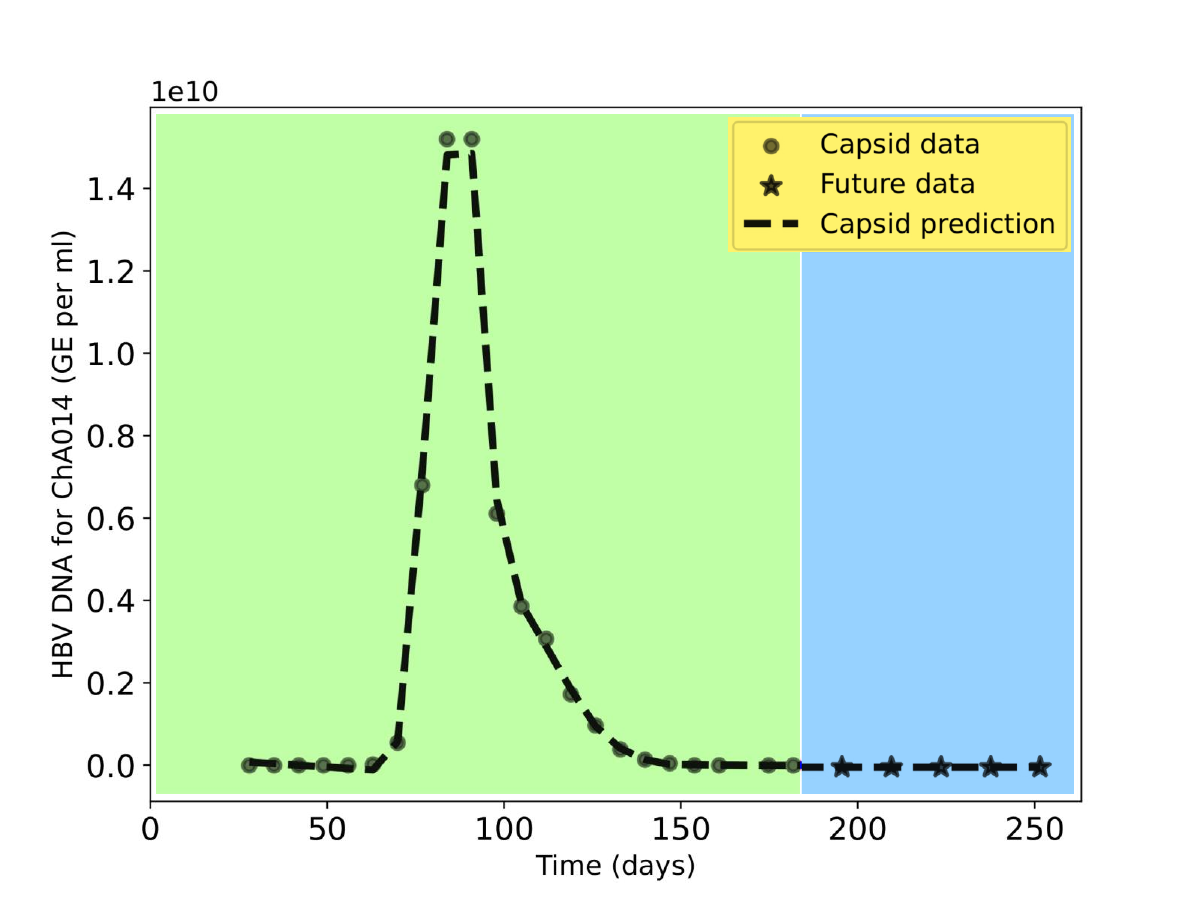}
    \caption{ChA014}
\end{subfigure}
\hfill
\begin{subfigure}[b]{0.3\textwidth}
    \centering
    \includegraphics[height=5.3cm, width=5.7cm]{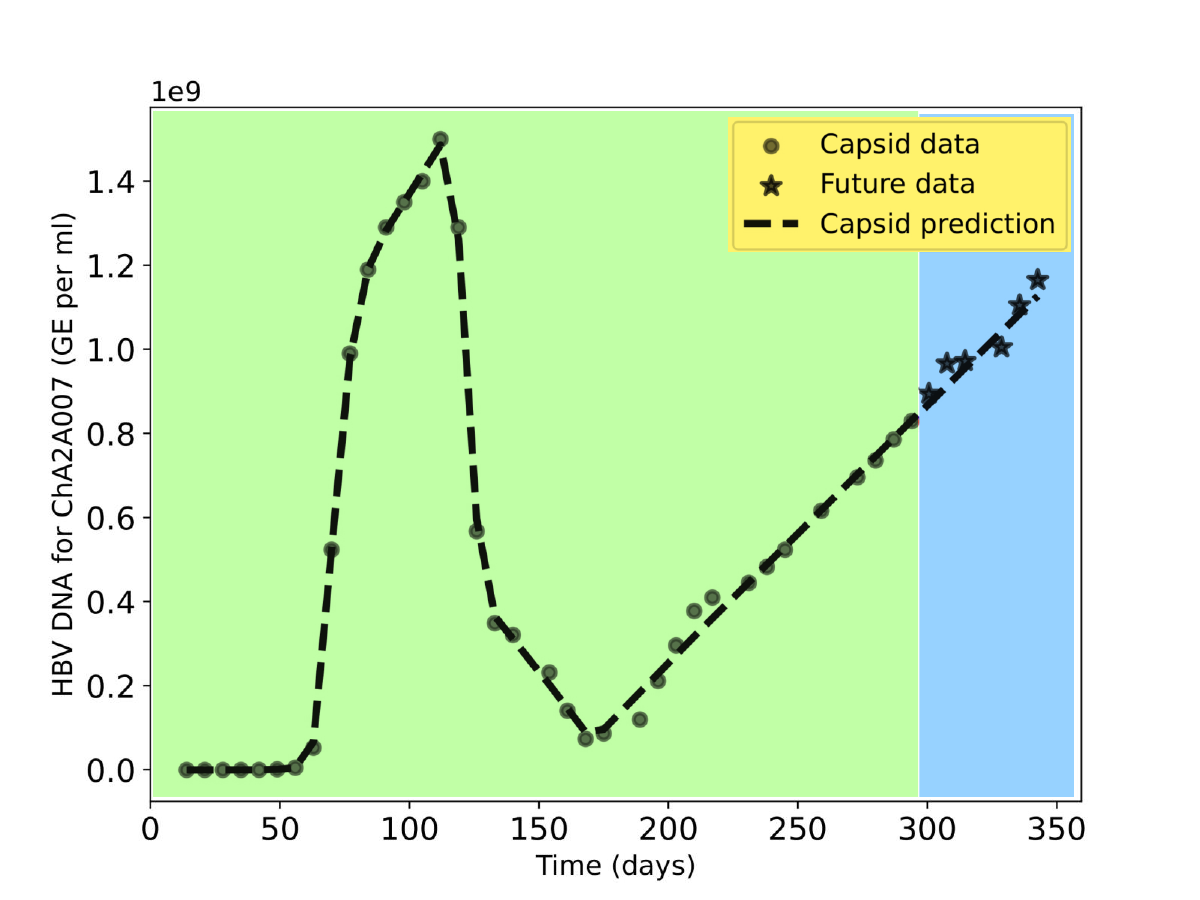}
    \caption{ChA2A007}
\end{subfigure}
\hfill
\begin{subfigure}[b]{0.3\textwidth}
    \centering
    \includegraphics[height=5.3cm, width=5.7cm]{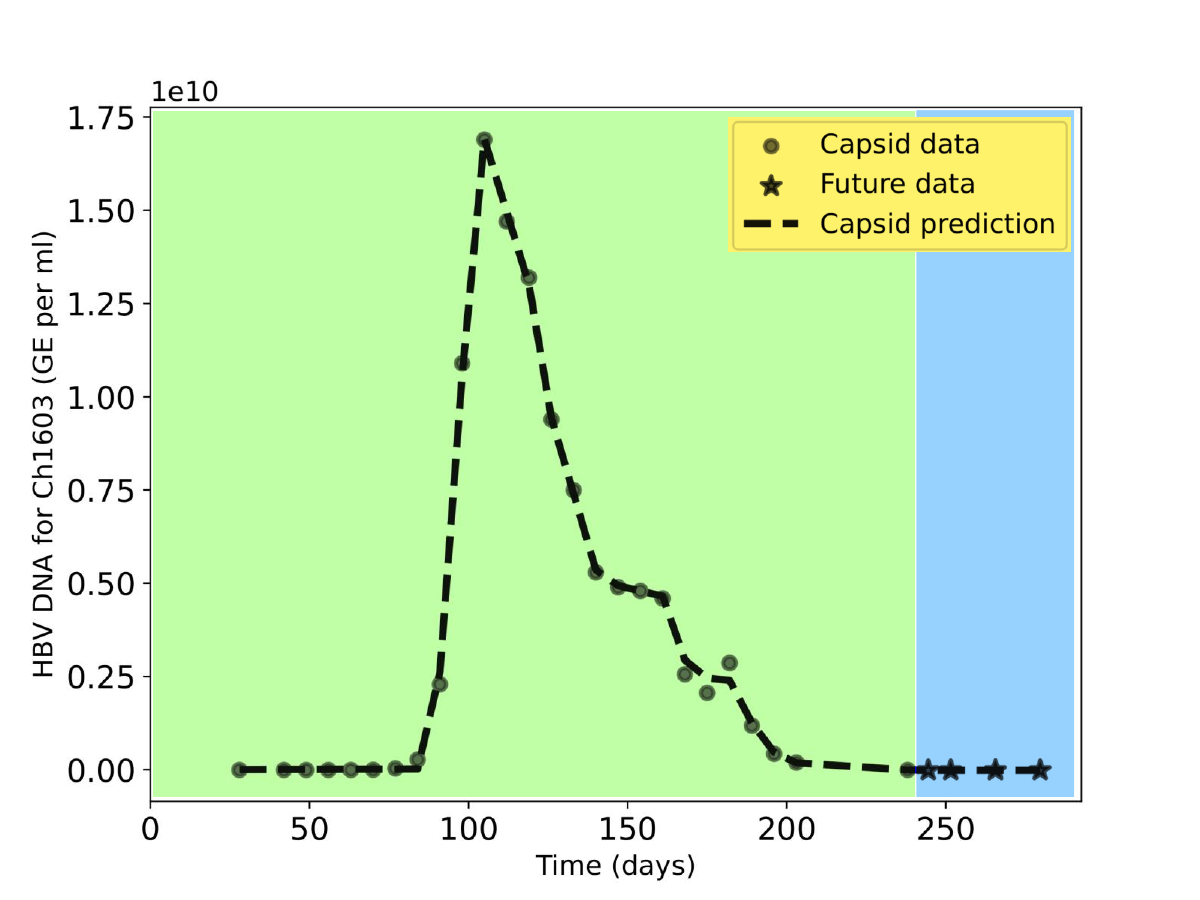}
    \caption{Ch1603}
\end{subfigure}    
\hfill
\begin{subfigure}[b]{0.3\textwidth}
    \centering
    \includegraphics[height=5.3cm, width=5.7cm]{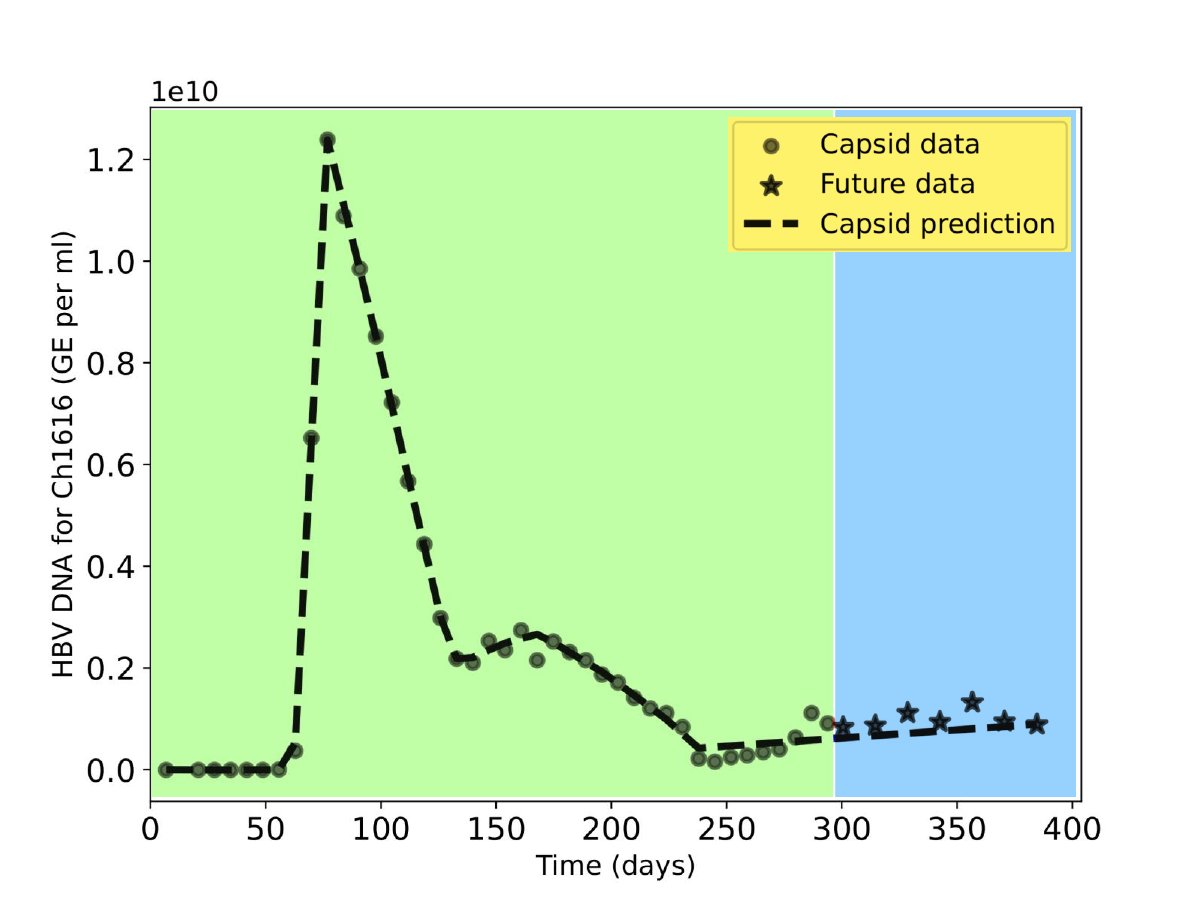}
    \caption{Ch1616}    
\end{subfigure}
\caption{Experimental validation of the model \eqref{main model} using experimental data from nine chimpanzees and future prediction. The data points in the green-shaded areas represent training capsid data, while those in the blue-shaded areas correspond to future capsid data.}
\label{Dynamic of HBV DNA for nine chimpanzees.}
\end{figure}
Simulations are performed for each of the nine chimpanzees with the same set-up as described above.  In Figure \ref{Dynamic of HBV DNA for nine chimpanzees.}, the experimental data and the solutions obtained from \textit{DINNs} are plotted for each chimpanzee. The training for each chimpanzee is performed using the data points marked with circles within the green-shaded region. Based on these training data, the network is able to predict the concentrations of HBV DNA-containing capsids both within and beyond the training interval. The predicted dynamics of capsids for each chimpanzee are represented by dashed curves (``\textbf{- -}'') in the subfigures of Figure~\ref{Dynamic of HBV DNA for nine chimpanzees.}.
The sky-blue shaded region indicates the testing window where the data points marked with asterisks represent the future capsid data that are kept for testing. These data points are used to assess how accurately the model can reproduce actual HBV DNA levels beyond the training window.
Upon comparing the solutions obtained by \textit{DINNs} with the experimental data, it is observed that the predictions for each chimpanzee agree well with the experimental data within and beyond the training window. \textit{DINNs} effectively captures the dynamics of both acute and chronic HBV infections. Moreover, in all nine experiments, the peaks of the capsid profiles are reliably identified. Therefore, it can be concluded that this methodology can also be used to predict the solution for unseen data.

\subsection{parameter estimation}
Out of the nine estimated parameters of the system \eqref{main model}, some of them demonstrate significant variability across chimpanzees, while others remain almost same. Specifically, \textit{DINNs} identifies that the values of the parameters $\lambda$, $\mu$, $k$, $\alpha$ and $c$ are nearly the same for all chimpanzees. This suggests that these parameters control basic viral and cellular processes that do not change much from one host to another.
In contrast, the remaining four parameters $a$, $\beta$, $\delta$ and $\gamma$ show significant variation across chimpanzees, indicating that these parameters can be considered sensitive to individual-specific biological factors, such as strength of the immune response, susceptibility, or differences in the initial concentration of HBV DNA across individuals. Consequently, these parameters are likely to influence host-specific outcomes, such as viral clearance versus viral persistence. In other words, they may play key roles in determining whether HBV infection becomes acute or develops into a chronic form.
In Table \ref{tab:chimp_parameters}, the values of these significant parameters are listed for each chimpanzee. For a better visualization, bar diagrams for all these parameters are displayed in the last column of Table \ref{tab:chimp_parameters}.

\begin{table}[h!]
    \centering
    \renewcommand{\arraystretch}{1.3}  
    \resizebox{\textwidth}{!}{%
    \begin{tabular}{|c|*{9}{>{\centering\arraybackslash}m{0.8cm}|}>{\centering\arraybackslash}m{8cm}|}
    \hline
                & \multicolumn{9}{c|}{\textbf{Chimpanzees}} & \\ \hline
    \textbf{P}  & \makecell{\rotatebox{90}{\textbf{ChA006}}} & 
                   \makecell{\rotatebox{90}{\textbf{ChA007}}} & 
                   \makecell{\rotatebox{90}{\textbf{Ch1622}}} & 
                   \makecell{\rotatebox{90}{\textbf{ChA3A005}}} & 
                   \makecell{\rotatebox{90}{\textbf{ChA2A007}}} & 
                   \makecell{\rotatebox{90}{\textbf{Ch1603}}} & 
                   \makecell{\rotatebox{90}{\textbf{Ch1616}}} & 
                   \makecell{\rotatebox{90}{\textbf{Ch1618}}} & 
                   \makecell{\rotatebox{90}{\textbf{ChA014}}} & \textbf{Bar diagram} \\ \hline
    
    \textbf{$a$} & \makecell{\rotatebox{90}{$251.068$}} & \makecell{\rotatebox{90}{$242.752$}} & 
                   \makecell{\rotatebox{90}{$217.663$}} & \makecell{\rotatebox{90}{$210.646$}} & 
                   \makecell{\rotatebox{90}{$210.030$}} & \makecell{\rotatebox{90}{$212.722$}} & 
                   \makecell{\rotatebox{90}{$202.911$}} & \makecell{\rotatebox{90}{$209.487$}} & 
                   \makecell{\rotatebox{90}{$213.720$}} & \includegraphics[width=0.97\linewidth]{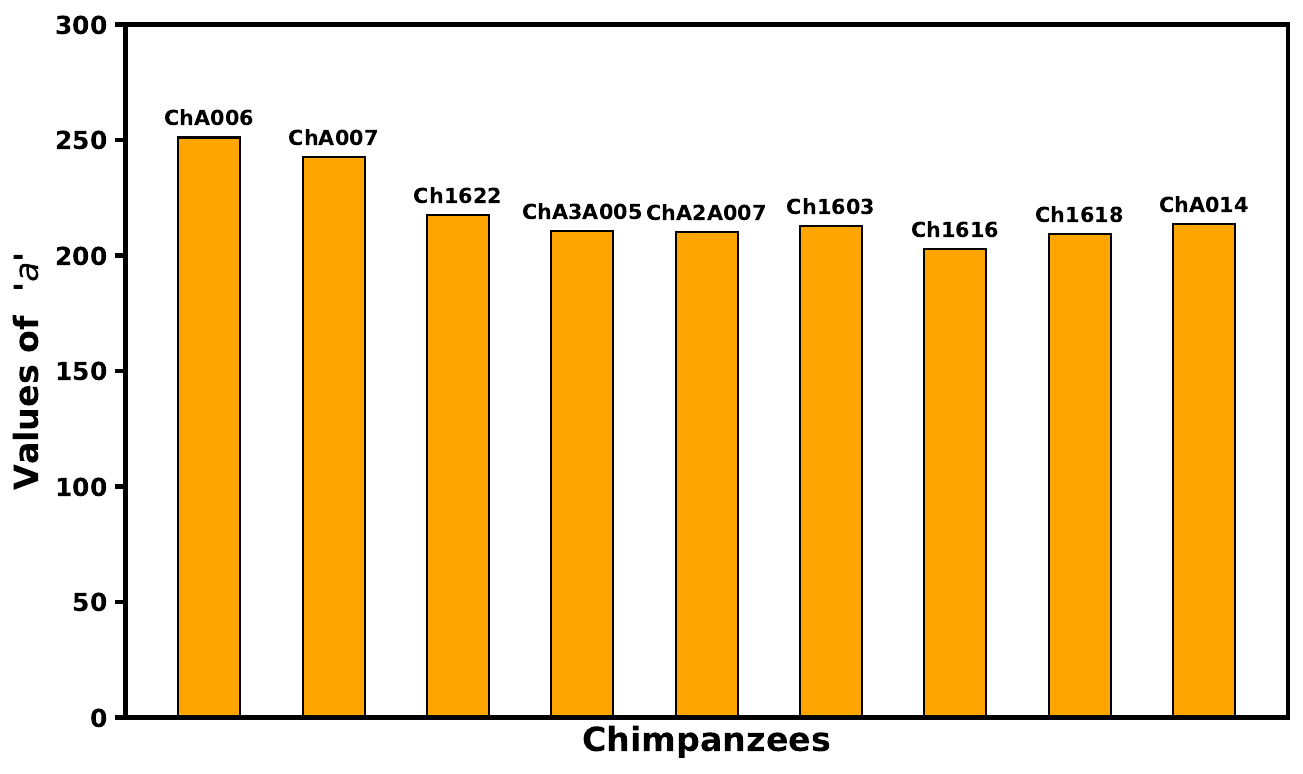} \\ \hline
    
    \textbf{$\beta$} & \makecell{\rotatebox{90}{$0.7768$}} & \makecell{\rotatebox{90}{$0.8289$}} & 
                        \makecell{\rotatebox{90}{$0.8375$}} & \makecell{\rotatebox{90}{$0.7650$}} & 
                        \makecell{\rotatebox{90}{$0.7542$}} & \makecell{\rotatebox{90}{$0.7423$}} & 
                        \makecell{\rotatebox{90}{$0.7822$}} & \makecell{\rotatebox{90}{$0.7555$}} & 
                        \makecell{\rotatebox{90}{$0.7581$}} & \includegraphics[width=0.97\linewidth]{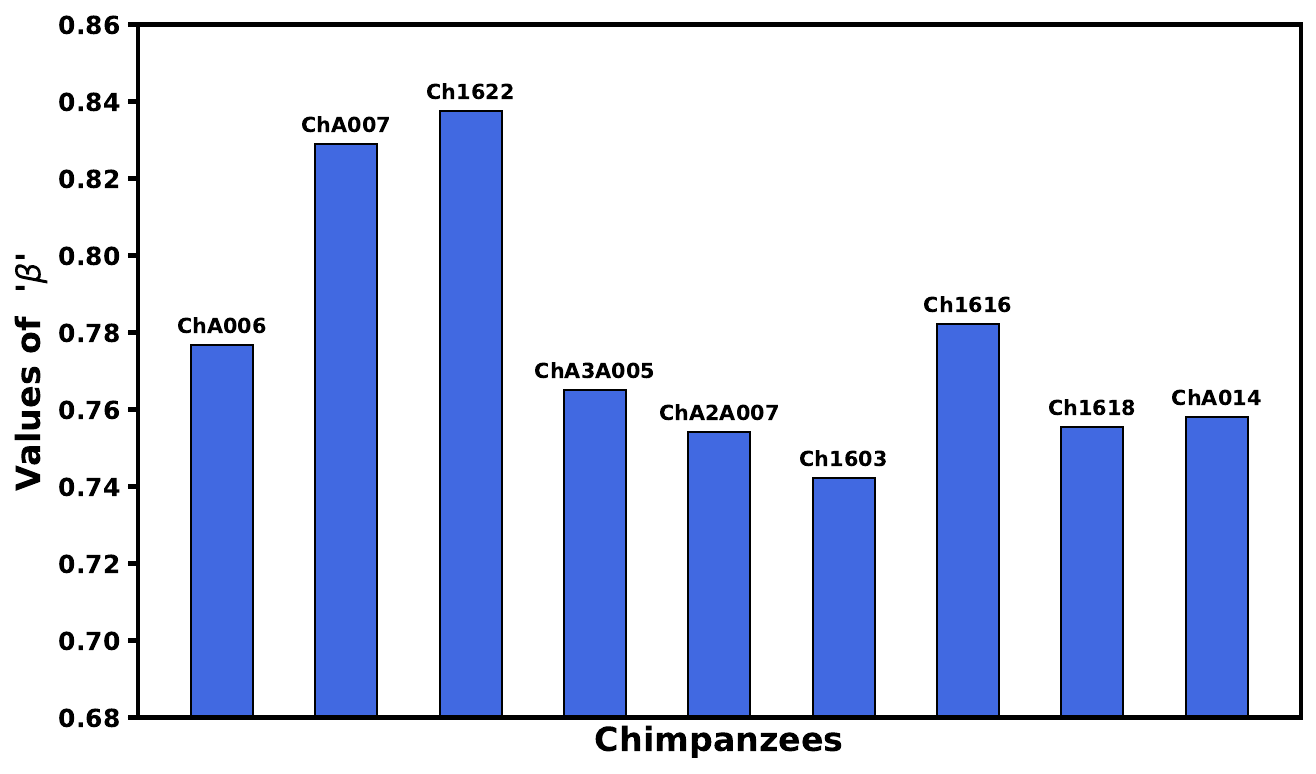} \\ \hline
    
    \textbf{$\delta$} & \makecell{\rotatebox{90}{$0.0788$}} & \makecell{\rotatebox{90}{$0.0813$}} & 
                        \makecell{\rotatebox{90}{$0.0816$}} & \makecell{\rotatebox{90}{$0.0784$}} & 
                        \makecell{\rotatebox{90}{$0.0778$}} & \makecell{\rotatebox{90}{$0.0773$}} & 
                        \makecell{\rotatebox{90}{$0.0792$}} & \makecell{\rotatebox{90}{$0.0779$}} & 
                        \makecell{\rotatebox{90}{$0.0780$}} & \includegraphics[width=0.97\linewidth]{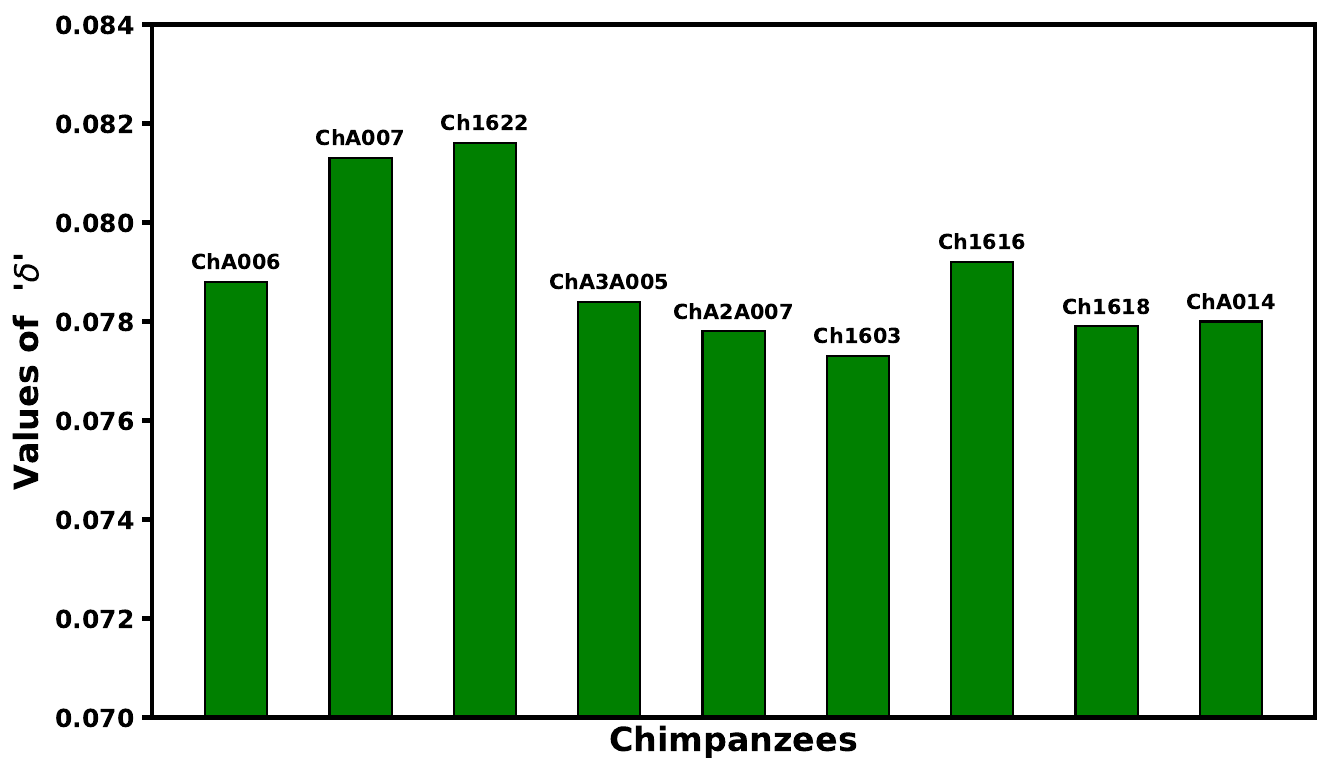} \\ \hline

    \textbf{$\gamma$} & \makecell{\rotatebox{90}{$1.0829$}} & \makecell{\rotatebox{90}{$1.0520$}} & 
                        \makecell{\rotatebox{90}{$1.0470$}} & \makecell{\rotatebox{90}{$1.0877$}} & 
                        \makecell{\rotatebox{90}{$1.0940$}} & \makecell{\rotatebox{90}{$1.1004$}} & 
                        \makecell{\rotatebox{90}{$1.0787$}} & \makecell{\rotatebox{90}{$1.0933$}} & 
                        \makecell{\rotatebox{90}{$1.0919$}} & \includegraphics[width=0.97\linewidth]{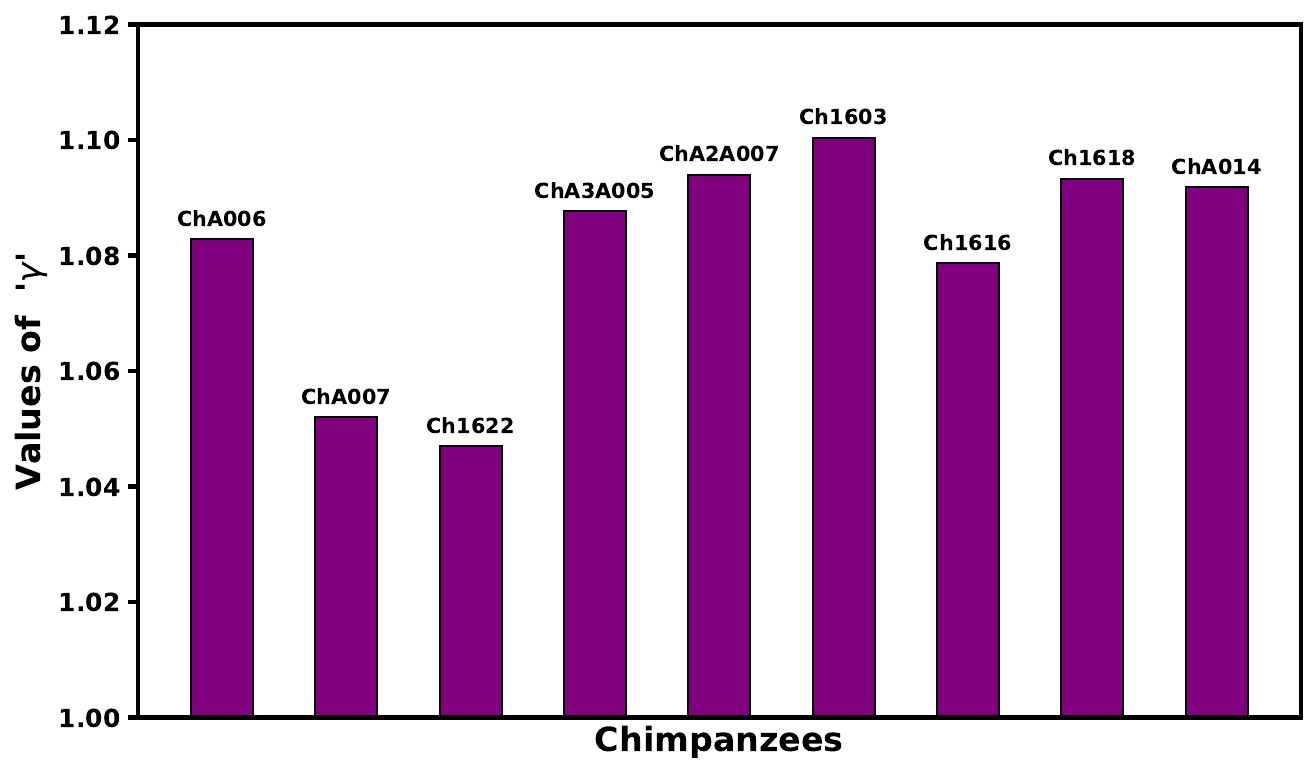} \\ \hline
    
    \end{tabular}%
    }
    \caption{Estimated values of the significant parameters for nine chimpanzees with corresponding bar diagrams for visualization.}
    \label{tab:chimp_parameters}
\end{table}

\clearpage
\section{Discussion and conclusions}
The inverse-forward framework is a traditional approach for forecasting infectious diseases. In this method, the parameters governing disease dynamics are first inferred from available data, and then used to generate future outcomes. In recent years, Physics-Informed Neural Networks (PINNs) have gained recognition as a powerful inverse-forward modeling framework, offering an alternative to traditional numerical methods. This approach capitalizes on the efficiency of neural networks to uncover the dynamics inherent in a system of differential equations by minimizing a loss function that combines both data loss and residual loss. Despite significant advances, standard PINNs often struggles to estimate the parameter values of a system from the available experimental data when the parameters have significant variations in the order of magnitudes relative to each other. Recently, \textit{Disease-Informed Neural Network (DINNs)}, a modified version of PINNs, have addressed these challenges and offers a robust technique for parameter estimation tasks, even when they differ by significant orders of magnitude. The key idea behind this technique lies in designing a feed-forward neural network that normalizes the input data and enables effective estimation of parameters associated with viral dynamics. Applying this advanced \textit{DINNs} approach, this study aims to investigate the complex dynamics of HBV infection. To this purpose, a recently proposed HBV infection dynamics model, as described by Sutradhar and Dalal \cite{r2024_sutradhacytoplasmic}, is considered.
In order to validate their model and estimate its parameters, they minimized the mean squared error between experimental data and predicted values of HBV DNA-containing capsids. However, relying solely on the data loss introduces a fundamental limitation: the model learns to fit the experimental data points but lacks enforcement of the underlying mechanistic constraints governing the system. This purely data-driven approach relies solely on interpolation and thus fails to ensure physically consistent dynamics outside the observed data range. 
Through \textit{DINNs}, we address these limitations by incorporating an additional residual loss term, which enforces the governing differential equations of the system \eqref{main model} during training. This inclusion ensures that the model not only fits the available data but also adheres to the fundamental physical principles underlying viral transmission and progression. By embedding domain knowledge into the learning process, \textit{DINNs} enhance model validation, improve parameter estimation, and provide biologically reliable predictions beyond the training interval.
In the context of HBV infection, 
the novelty of these \textit{DINNs} approach is two-fold: (i) the determination of the concentrations of uninfected hepatocytes, infected hepatocytes, HBV DNA-containing capsids, and virions based on the temporal information, and (ii) the estimation of unknown model parameters by employing a fast and comprehensive inverse framework.

A comprehensive summary of the numerical results by \textit{DINNs} is presented in two separate scenarios. Initially, experiments were conducted over a time span of 500 days, varying parameter ranges, noise levels, data sample sizes, network architecture, and learning rate using numerically synthesized data. In nearly all cases, \textit{DINNs} are capable of fitting the data well and accurately capturing the dynamics of the model compartments. The results obtained in Table \ref{tab:optimal_parameters} - \ref{tab:Performance of DINNs for sample sizes} show remarkable precision, particularly in retrieving the model parameters. 
The numerical experiments shown in Subsection \ref{Impacts of noise on parameter estimation} demonstrate that this method works well, even in the presence of noisy datasets.
To further improve the realism of viral dynamics and enhance the robustness of model predictions, the proposed model was validated using experimental data of nine chimpanzees.  
Despite the data scarcity, the model outcomes obtained from \textit{DINNs} for each chimpanzee demonstrate a good agreement to the experimental data. The model solutions, shown in Figure \ref{Dynamic of HBV DNA for nine chimpanzees.}, reveal the capability of this method in identifying the peaks of the capsid profiles. Since intracellular capsids are precursors to mature virions, these peaks mark the point of highest intracellular viral replication and consequently represent the most critical phase of HBV infection. So, accurately identifying these peaks can help in optimizing the timing of antiviral interventions. 
 Moreover, a key advantage of \textit{DINNs}, compared to traditional numerical methods, is their capacity to provide reliable forecasts beyond the observation interval.
This capacity arises from the fact that \textit{DINNs} incorporate the residuals of the ODE system \eqref{main model} into the loss function during training, which helps the model to track the underlying dynamics and current trend of the infection. As shown in Figure \ref{Dynamic of HBV DNA for nine chimpanzees.}, this approach is crucial in capturing the dynamics of capsids for future cases. 
It is also observed that \textit{DINNs} are able to effectively identify the most significant parameters ($a$, $\beta$, $\delta$ and $\gamma$) in HBV dynamics when only the experimental data for the HBV capsids are used. These four parameters exhibit substantial variations across chimpanzees and may be considered the most critical in determining whether an infection is cleared or becomes chronic.

In conclusion, the application of Physics-Informed machine learning to Hepatitis B virus infection presents a powerful, flexible, and biologically informed framework for monitoring disease progression. By reliably estimating the underlying parameters and fitting the experimental data effectively, this method ensures trustworthy forecasts of HBV dynamics. These findings may provide valuable insights into viral infection processes, which are essential for guiding public health strategies and improving the precision of clinical interventions.


\subsection*{Acknowledgments}
The first author would like to acknowledge the financial support obtained from the University of Grants Commission, Government of India (File No: 211610121834). All the authors thank the research facilities received from the Department of Mathematics, Indian Institute of Technology Guwahati, India.
\subsection*{Data availability statement}
 Data sharing is not applicable. 
 \subsection*{Author contributions}
\noindent Bikram Das: Visualization, Validation, Conceptualization, Data collection, Methodology, Writing-original draft.

 \noindent Rupchand Sutradhar: Conceptualization, Investigation, Methodology, Writing-original draft.

 \noindent D C Dalal: Conceptualization, Formal analysis, Review and editing-original draft, Supervision.

 \subsection*{Conflict of interest}
The authors declare no potential conflict of interests.

\subsection*{Ethical approval}
This study did not conduct any experiments on human participants or animals. All data used were either publicly available or numerically synthesized.


\end{document}